# Unsupervised Online Bayesian Autonomic Happy Internet-of-Things Management

Rossi Kamal[a], Choong Seon Hong[a,*]

[a]*Kyung Hee University*

## Abstract

In *Happy IoT*, the revenue of service providers synchronizes to the unobservable and dynamic usage-contexts (e.g. emotion, environmental information, etc.) of Smart-device users. Hence, the usage-context-estimation from the unreliable Smart-device sensed data is justified as an unsupervised and non-linear optimization problem. Accordingly, *Autonomic Happy IoT Management* is aimed at attracting initial user-groups based on the common interests (i.e. *recruitment*), then uncovering their latent usage-contexts from unreliable sensed data (i.e. *revenue-renewal*) and synchronizing to usage-context dynamics (i.e. *stochastic monetization*). In this context, we have proposed an unsupervised online Bayesian mechanism, namely *Whiz* (Greek word, meaning Smart), in which, (a) once latent user-groups are initialized (i.e *measurement model*), (b) usage-context is iteratively estimated from the unreliable sensed data (i.e. *learning model*), (c) followed by online filtering of Bayesian knowledge about usage-context (i.e. *filtering model*). Finally, we have proposed an Expectation Maximization (EM)-based iterative algorithm *Whiz*, which facilitates *Happy IoT* by solving (a) *recruitment*, (b) *revenue-renewal* and (c) *stochastic- monetization* problems with (a) *measurement*, (b) *learning*, and (c) *filtering* models, respectively. Through theoretical analysis and synthetic dataset results, *Whiz* is justified as the optimal solution of *Online Stochastic Fact-Finding* for not only estimating the likelihood of usage-context (i.e. fact), but also quantifying its global performance with Bayesian knowledge, by outperforming counterpart *revenue-renewal* model (i.e. Tube) and *fact-finding* schemes (i.e. voting, pagerank, TruthFinder, Bayesian-based, EM-based), in terms of the estimation accuracy, and convergence. Moreover, e-mail survey dataset is utilized for ground-truth selection in prototype development on Android and web platforms in *Happy IoT* scenario, followed by *Whiz*-driven approximation of dynamic usage-contexts and filtering of social trends through YouTube, Cenceme and Twitter datasets, respectively.

---

[*]Choong Seon Hong is the corresponding author
   *Email addresses:* rossi@khu.ac.kr (Rossi Kamal), cshong@khu.ac.kr (Choong Seon Hong)

| $IF(x)$ | $COI(\theta)$ | $PlayerA$ | $Player\ B$ |
|---|---|---|---|
| Observation | Parameter | Service provider | User |

Table 1: Terminologies used in *Autonomic Happy IoT Management*

1. Introduction

*1.1. Motivation*

IoT has proliferated as the digital presence of various things or objects, such as sensors, actuators and Smart-devices[1][2]. The penetration of smart-devices and the increasing demand of personalized services have attracted numerous service providers, who want to monetize from Smart-device user-groups based on their common interests, such as emotion (e.g in personal search [3], well-being [4] or personalized shopping applications [5]); location (e.g in traffic-signal-detection [6][7], transit tracking [8][9], localization [10][11], sociability detection [12], consumer-care [13][14][15], media-sharing [16] and transportation [17]) and environmental information(e.g in environmental report generation [18][19][20], rare event detection [21][22][23]). In this context, service providers seek for enhanced revenue though different incentivization/motivational steps, such as monetary support [24][25][26], social reputation [27][28][29], quality-of-information assurance [30][31] or even enhanced consumer care [32][33][34]. However, it necessitates the continual regulation of usage-dynamics through conventional popularity-measurement-tools (e.g. download-count, user-rating, usage-duration, etc.) and even, IoT-renovated service-usage measurement tools[35][36][37].

*1.2. Happy IoT*

Let us illustrate *Happy IoT* (Definition. 1) (Fig. 1) with preliminary notation convention and terminologies (Table. I).

*Notation Convention:* Scalar quantity ($A$) is denoted by Italic; vector quantity($a$) is denoted by lowercase boldface; a matrix quantity($A$) is denoted by uppercase boldface. The matrix transpose is denoted by a superscript ($A^T$). The n-th row and m-th column element of the matrix $A$ is represented by $\{A\}_{n,m}$. $I_N$ denotes the identity matrix of size $N \times N$. $0_{N \times M}$ is a matrix of zeros. $\|.\|$ denotes the norm. $|.|$ denotes the modulus. $abs(.)$ denotes the absolute value. $\Omega$ is the observation(Incentive Filter) space and $\Theta$ is the parameter (COI) space. $E[.]$ denotes the expectation operator with repect to a density probability function explicitly given by a subscript.

We envision a *Happy IoT* (Fig. 1) (Definition. 1) (Table. 1), in which the revenue of service providers (denoted as $PlayerA$) synchronizes to the user-experience of Smart-device users (denoted as $PlayerB$). In this context, $PlayerA$ deploys various measurement tools (denoted as Incentive Filter $IF(x)$)(e.g usage-duration, usage-frequency, real-time feedback and service execution, etc.)[36][35] to infer usage-context (denoted as Choice-of-Interest $COI(\theta)$) (e.g. emotion, weather, location and time, etc.) of $PlayerB$. Moreover, $PlayerA$ often utilizes various sensors (e.g. traffic monitoring, survey) to collect real-time/offline feedback of $PlayerB$ to regulate whether their business goals (e.g. popularity, financial profit) are met (Fig. 2).

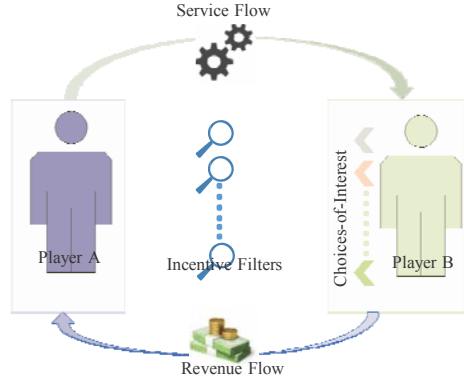

Figure 1: Players in *Happy IoT*

**Definition 1.** *Happy IoT*: $PlayerA$ is Happy about the revenue on any personalized service, when he/she is satisfied with revenue observed with usage-dynamics through $IF(x)$ and it is denoted by $P_1(x) = 1$. Similarly, $PlayerB$ is Happy about personalized service, when he/she is satisfied with the $COI(\theta)$ and it is denoted by $P_2(\theta) = 1$. Therefore, Happy IoT means, $PlayerA$ and $PlayerB$ are Happy, when $PlayerA$ is able to estimate $PlayerB$'s $COI(\theta)$ from sensed data of $IF(x)$.

*1.3. Challenges of Autonomic Happy IoT Management*

$PlayerA$ seems to care about the quality of submitted information[30] (e.g. large corpus of incomplete data[38], recovery of random samples[39] of Smart-device data) and even about the time dependency between usage-contexts, behavior of $PlayerB$. An unsupervised mechanism[40] is essential to discover latent usage-contexts from the unreliable sensed data. Because, Smart-device traffic is often impaired/unreliable because of poor sensor quality, lack of calibration techniques, absence of human attention, and even his/her intention to deceive[41][42][43]. Moreover, it is essential to synchronize business goals (e.g. popularity, income) to the time dependency between current or future behavior of $PlayerB$ in different contexts[25] and even critical events[18].

In this context, Emotionsense[44] uses Gaussian mixture model to infer emotions, as well as activities, verbal and proximity interactions among social groups. However, Tube[45] enables monitoring Smart-device usage patterns to estimate Smart-device users' willingness to shift their usage in exchange for a monetary discount. Voting[46], pagerank[47], Bayesian[43], maximum likelihood[41] and Cramer-Rao lower bound[42] have been devised to estimate latent contexts in information network. However, these proposals suffer from deficiencies like the absence of credible up-to-date data, sensitivity to initialization[43], local likelihood estimation[41] and the absence of global quantification[42], respectively.

An autonomic approach is recognized as an enabling technology to adapt to dynamic environmental contexts in network and service management[48]. IBM's MAPE (Monitor, Analyze, Plan and Execute)(Fig. 3 (a))-based autonomic approach, is frequently utilized to *M*onitor and *A*nalyze Smart device traffic and consequently, *P*lan, *E*xecute dynamic context-specific network/service management policies[49]. Consequently, autonomic management of *Happy IoT* requires $PlayerA$ to synchronize his/her autonomous service recommendation policies according to usage-dynamics (e.g. real-time feedback, service visualization) of $PlayerB$ (Fig. 2). Therefore, the major challenge resides in the way it integrates Bayesian knowledge to facilitate synchronization in unsupervised usage-context (i.e $COI(\theta)$) estimation.

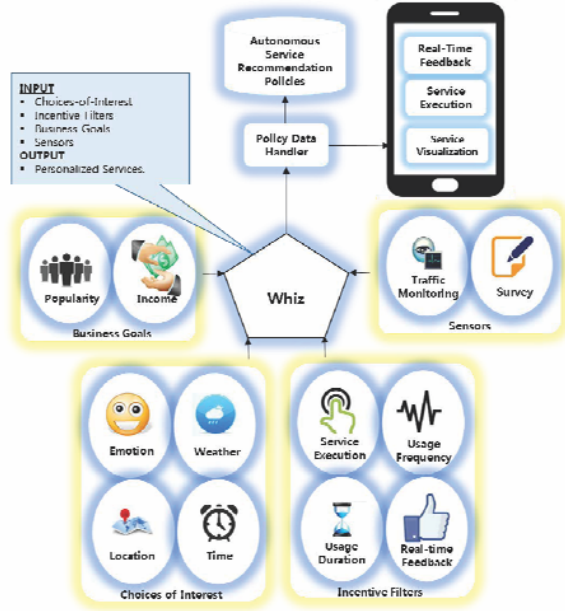

Figure 2: Various components of $Happy\ IoT$

*1.4. Problem Formulation*

We now formulate *Autonomic Happy IoT Management* problem (Definition. 2)(Fig. 3(b)).

Definition 2. $Autonomic\ Happy\ IoT\ Management$: Autonomic Happy IoT Management enables $PlayerA$ *to renew iteratively revenue-making process according to the dynamic* $COI(\theta)$ *of* $PlayerB$. *In this context, (a) at first user-groups are recruited from the distribution of* $COI(\theta)$; *(b) then* $COI(\theta)$ *is estimated from the sensed data of* $IF(x)$, *then; (c) Bayesian knowledge is used to adapt to* $COI(\theta)$, *which might be random with a prior probability distribution. Thus, Autonomic Happy IoT Management problem is decomposed into following three sub-problems.*

*1.4.1. Recruitment Problem*

It addresses the attraction of common user-groups. Hence, *an optimal solution is to be devised, which attracts user-groups with common interests* (Lemma. 1).

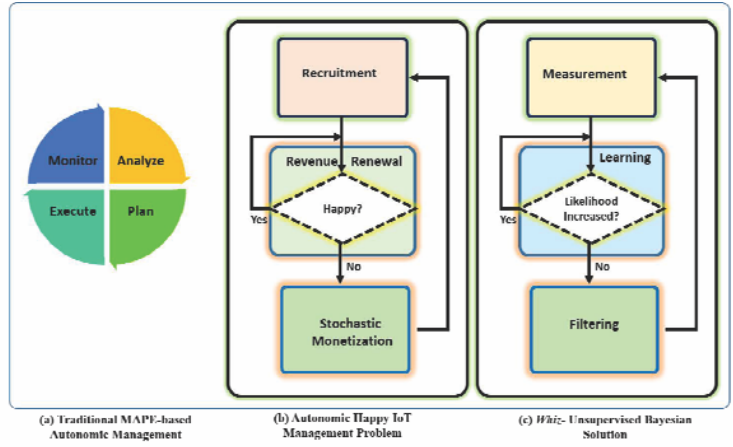

Figure 3: (a)Traditional MAPE-based autonomic management problem, (b) devised *Autonomic Happy IoT management problem*, (c) proposed *unsupervised Bayesian solution* Whiz

*1.4.2. Revenue-renewal Problem*

It addresses *the estimation of $COI(\theta)$ from the sensed data of $IF(x)$, especially if $IF(x)$ does not have enough information about $COI(\theta)$. Hence, an EM-based iterative algorithm is to be devised, which should converge to an optimal solution*.

*1.4.3. Stochastic-Monetization Problem*

It addresses the iterative checking of the minimum bound on mean-square-error (MSE) of $COI(\theta)$-estimation from the observed $IF(x)$, when $COI(\theta)$ is random with a prior probability distribution. Hence, *the optimal solution should not only estimate COI but also quantify its global (i.e. time variant[50]) performance with Bayesian knowledge*.

*1.5. Unsupervised Online Bayesian Mechanism*

Therefore, we have proposed (Section 2) an unsupervised online Bayesian mechanism *Whiz* [51], which enables $PlayerA$ to iteratively estimate the dynamic usage-contexts from the unreliable sensed data of $PlayerB$. Hence, *Whiz* solves (a) *recruitment*, (b) *revenue-renewal*, and (c) *stochastic monetization* problems, by (a) *measurement*, (b) *learning*, and (c) *filtering* models (Fig. 3(c)) though a final algorithm (i.e *Whiz*) as follows

*1.5.1. Measurement Model*

It initializes user-groups with common interests through $K-means$ clustering[52]. Accordingly, *measurement model works as an optimal solution of recruitment model (Lemma. 4)*.

*1.5.2. Learning Model*

As *revenue-renewal problem* is amenable to general EM[53] (Lemma. 3), learning model facilitates the estimation of the likelihood of $COI(\theta)$ from the incomplete sensed data (Lemma. 8) with common interests through *Whiz*-EM (Lemma 9-13). Accordingly, *learning model solves* revenue-renewal problem *having latent user-group as a mixture component of the Gaussian mixture model (Theorem. 4)*

*1.5.3. Filtering Model*

It facilitates synchronization to the frequent changes in common interests through the global quantification (Theorem. 5) of the likelihood of $COI(\theta)$-estimation with Bayesian online filtering. Accordingly, *filtering model works as an optimal solution of stochastic monetization (Theorem. 6)*.

*1.5.4. Whiz Algorithm*

In *Whiz* algorithm, (a) once latent user-groups are initialized (*measurement model*), (b) their latent $COI(\theta)$ is iteratively estimated from the sensed data of *IF* (*learning model*), (c) followed by the Bayesian quantification to adapt to the changes on prior knowledge about $COI(\theta)$ (*filtering model*).

*1.6. Theoretical Analysis*

The theoretical analysis (Section 2) gives us the key findings

(a) *Revenue-renewal* is proven as an unsupervised (Lemma. 2) and non-linear optimization (Theorem. 2) problem, followed by the technical validation how it is amenable to finite mixture model EM (Lemma. 3). Accordingly, finite mixture model with latent variable (representing latent user-group) is justified to formulate estimable $COI(\theta)$ (Theorem. 1), which is necessary in EM-based unsupervised online Bayesian scheme.

(b) *Stochastic monetization* is proven as a *Online Stochastic Fact-Finding* (Definition. 4) problem (Theorem. 3), preceded by the technical validation of latest fact-finding research[42] as a deterministic problem (Lemma. 7).

**Definition 3.** Online Stochastic Fact-finding*: Online Stochastic fact-finding is the procedure of estimating fact/parameter $COI(\theta)$ from the sensed data of observed IF x), where as soon as k-th sensed data from IF ($(x_k)$ is received, Online Bayesian bounds[?] are used as a lower bound of the global MSE in $COI(\theta)$-estimation. Online Bayesian filtering estimate $COI(\theta_k)$ by using current and previously sensed data ($x_1^K = [x_1, x_2,....x_k]^T$) of IF.*

*Online Stochastic Fact-finding*, in contrast to state-of-the-art fact-finding[46][47][54][43][41][42], which often lack convergence-guarantee and appropriate objective functions to optimize, is justified as an unsupevised (Lemma. 2) and non-linear optimization (Theorem. 2) problem.

(c) *Whiz* is proven as an optimal solution of *Online Stochastic fact-finding* (Theorem. 6), followed by the experimental proof of faster convergence than existing fact-finders (Section IV). In addition to original *EM* [53], which iteratively estimates parameter $COI(\theta)$ from the sensed data, *Whiz*-EM quantifies globally the confidence in estimation results based on Bayesian Cramer-Rao lower bound[50] from synchronization theory (Theorem. 5).

*1.7. Empiricial Results*

Numerical experiments (Section 3) on synthetic datasets justify that, with the increase in the number of *IF* or *COI* or *COI* per *IF* and the ratio of satisfactory *COI*, demand frequency of *IF*, *Whiz* outperforms counterpart deterministic fact-finding schemes (i.e. voting[46], pagerank[47], Bayesian[43], maximum likelihood[41]-based, Cramer-Rao lower bound[42]-based) in terms of the estimation accuracy and convergence on synchronization to the dynamic usage-contexts. Moreover, utilizing an Email survey datasetpsychic, prototypes are developed on Android and Web platforms in a *Happy IoT* scenario of personalized service recommendation for work-going persons. Finally, a YouTube[55] and a Twitter[56] datasets are trained with Whiz to approximate dynamic usage-contexts in media traffic and to filter rumor trend in social networks, respectively.

*1.8. Organization*

The paper is organized as follows-(a) Unsupervised Bayesian scheme is presented in section 2, (b) Theoretical analysis is presented in Section 3, (c) Empirical results are presented in Section 4, (d) Related work and Conclusion are presented in Section 5 and 6, respectively.

2. Unsupservised Online Bayesian Scheme

In this section, we have proposed an unsupervised online Bayesian algorithm, *Whiz* (Algorithm. 1), that enables service providers to iteratively estimate $COI(\theta)$ of Smart-device users from the sensed data by quantifying the global likelihood estimation with the advent of prior knowledge. Inputs of the algorithm are $COI(\theta)$ and IF($x$) and the output is the likelihood of $COI(\theta)$ from the sensed data. The algorithm works as follows- at first latent user-groups are initialized with measurement model. Then, $COI(\theta)$ is iteratively estimated from the sensed data until the likelihood increases in learning model. This iteration involves two steps, namely E-step and M-step, respectively. At $E$-step, latent user-group is inferred, which is followed by the $M$-step, when preference, discrepancy and mixture policy are upgraded. However, whenever the likelihood of $COI(\theta)$-estimation starts to decrease, the EM-iteration is terminated by filtering model. The iteration inside filtering models starts by checking the presence of prior knowledge about $COI(\theta)$. The absence of the any prior knowledge drives the Cramer-Rao lower bound to quantify the estimator. Whenever the prior knowledge about $COI(\theta)$ changes frequently, online filtering is used to quantify the likelihood estimator (Definition. 4).

3. Theoretical Analysis

In this section, we validate *Autonomic Happy IoT Management* problem and its sub-problems, followed by the justification of the proposed *Whiz* and its different models.

*Recruitment* is intended to attract user-groups with common interests. This is justified as a multivariate Normal distribution, having COI($\theta$) as multi-variables (Lemma. 1).

Lemma 1. Recruitment *is a multivariate Normal distribution problem*

---
**Algorithm 1:** Whiz Algorithm
---
  **Data:** $IF(x)$, $COI(\theta)$
  **Result:** Likelihood of $COI(\theta)$ from observed $IF(x)$
1 **begin** *Learning*
2    **begin** *Measurement*
3       **while** *(max $\sum_{n=1}^{N} \sum_{k=1}^{K} \{R\}_{n,k} ||x_n - \mu_k||$ )* **do**
4          (a) Assign every user($x_n$) to user-group ($\mu_k$)
            (b) Re-calculate preference($\mu$) of each user-group.
5       **end**
6    **end**
7    **begin** *Learning*
8       **while** *(max($\sum_{n=1}^{N} \ln \{\sum_{k=1}^{K} \pi_k N(x_n|\mu_k, \Sigma_k)\}$)* ) **do**
9          (a) Choose the user-group that has given accurate prediction in the recent past (Lemma. 10).
10         (b) Update preference (Lemma. 11), discrepancy (Lemma. 12), mixture strategies (Lemma. 13) for each user-group.
11       **end**
12    **end**
13    **begin** *Filtering*
14       **while** *(1)* **do**
15          **if** *Player A has no prior assumption about Player B's $COI(\theta)$,* **then**
16             Quantify locally the likelihood of $COI(\theta)$-estimation(Lemma. 14).
17          **end**
18          **else if** *Player A has prior assumption about Player B's COI-vector($\theta$), that is $P(\theta)$* **then**
19             **if** *Bayesian knowledge changes frequently* **then**
20                 Quantify $COI(\theta)$-estimation Online (Definition. 4)
21             **end**
22          **end**
23       **end**
24    **end**
25 **end**

*Proof.* Recruitment necessitates the attraction of user-groups with common interests. It is also significant to derive the variance of interests of any user-group from the most common user-group. Hence, given sensed data from the observed IF($x$), *recruitment* addresses (a) the distribution of COI($\theta$), (b) the correlation of COI($\theta$), (c) the uncertainty of one COI($\theta$) with respect to correlated COI($\theta$). Therefore, assuming COI($\theta$) as multivariables, *recruitment* becomes a multivariate Gaussian distribution problem as follows.

$$N(x|\mu, \Sigma) = \frac{e^{0.5(x-\mu)^T (\Sigma)^{-1}(x-\mu)}}{(2\pi)^{(n)/2} \sqrt{\Sigma}} \tag{1}$$

where *preference matrix* ($\mu$) calculates the most common user-group for COI($\theta$) and *discrepancy matrix*($\Sigma$) calculates the variances of user-group, from the common user-group for COI($\theta$).

□

However, *revenue-renewal* is intended to derive the usage-dynamics from the unreliable Smart-device data. Hence, the estimation of COI($\theta$) from the Smart-device data is justified as unsupervised (Lemma. 2) and non-linear optimization (Theorem. 2) problem, preceded and followed by justification of its amenability to finite mixture model (Theorem. 1) and EM (Lemma. 3), respectively.

Lemma 2. Revenue-renewal is an unsupervised learning problem

*Proof.* Revenue-renewal necessitates COI($\theta$) estimation from Smart-device data, which is often impaired because of poor sensor quality, lack of calibration or human attention and even his/her intention to deceive[41][42][43]. Hence, it is required to estimate most likely $COI(\theta)$ out of all-chances, that could have produced the sensed data of $IF(x)$. Therefore, inferring these $COI(\theta)$ and identifying which $COI(\theta)$ have produced these sensed data of $IF(x)$, lead to the creation of user-group from the set of sensed data. Hence, learning of $COI(\theta)$ from the sensed data of $IF(x)$ is an unsupervised learning problem. □

Theorem 1. Revenue-renewal is amenable to finite mixture model

*Proof. Revenue-renewal* is amenable to finite mixture model with latent variables, such that, $IF(x)$ is determined from one of $k$ latent $COI(\theta)$, which are with varying different probabilities.

Assuming $j$-dimensional $IF(x)$ follows a $k$-component finite mixture distribution, we obtain

$$p(x_j|\theta) = \sum_{k=1} \alpha_k p(x_j|\theta_k) \tag{2}$$

subject to

$$\sum_{k=1} \alpha_k = 1 \tag{3}$$

where $\alpha_k$ is the mixing coefficient and $\theta_k$ is the set of $COI$ parameters of the $k$-th mixture component $p(x_j|\theta_k)$. Therefore, $\theta = \{\alpha_1, \theta_1, \alpha_2, \theta_2, ..., \alpha_k, \theta_k\}$ be the complete set of parameters, those define this finite mixture model of revenue-renewal problem. □

Theorem 2. Revenue-renewal is a non-linear optimization problem

*Proof.* Given a set of $j$ independent and identically distributed samples of IF(x), the log-likelihood corresponding to a $k$-component mixture is

$$logp(x|\theta) = \sum_{j=1} log \sum_{k=1} \alpha_k p(x_j|\theta_k) \tag{4}$$

Hence, the objective function of estimating $COI(\theta)$ from the sensed data of $IF(x)$ is maximizing its log-likelihood criterion. Since, maximizing the logarithm of any likelihood is a non-linear optimization problem, revenue-renewal problem (maximizing log likelihood estimator of $COI(\theta)$) is turned to a non-linear optimization problem as follows

$$\hat{\theta}_{ML} = argmax_\theta \{logp(x|\theta)\} \tag{5}$$

□

Lemma 3. *Revenue-renewal is amenable to EM*

*Proof.* Since $\hat{\theta}_{ML}$ in Equation. 5 is not analytically tractable, EM is a natural choice to iteratively find the maximum likelihood solution of $\hat{\theta}_{ML}$ of revenue-renewal problem. Hence, revenue-renewal is amenable to EM. □

*Stochastic monetization* is intended to synchronize to the continual changes on common interests. Hence, *stochastic monetization* problem is formulated to quantify the global performance of the likelihood of COI estimation in Bayesian settings. Hence, let us validate how *Autonomic Happy IoT Management* lays out the foundation of *Online Stochastic Fact-finding* (Theorem. 3).

Theorem 3. Autonomic Happy IoT Management lays out the theoretical foundation of Online Stochastic fact-finding

*Proof.* Autonomic Happy IoT Management necessitates synchronization to the dynamic usage-contexts, which is possible with iterative quantification of global performance of the likelihood of COI($\theta$) estimation. This is only achievable with fact-finding with Bayesian bounds, which perform as a lower bound of the global MSE for COI ($\theta$)-estimation[50][57][58]

$$\text{MSE}_G = \int_\Phi \int_\Omega (\hat{\theta}(x) - \theta_0)(\hat{\theta}(x) - \theta_0)^T)^T p(x, \theta) dx d\theta \tag{6}$$

where $\theta\epsilon\Theta$ is random COI with a priori probability density function $p(\theta) = (p(x, \theta)/p(x/\theta))$ and $p(x, \theta)$ is the joint probability function of the sensed data of observed IF($x$) and COI ($\theta$). Hence, Autonomic Happy IoT Management lays out the theoretical foundation of Online Stochastic fact-finding. □

Let us justify *Measurement* model as an optimal solution of recruitment model, so that user-groups with most common interest is attracted (Lemma. 4).

Lemma 4. Measurement model works as an optimal solution of recruitment model

*Proof.* Measurement model initializes user-groups with common interest through K-means clustering[52]-based for (a) re-assigning sensed data to user-groups, and (b) re-computing the preference of user-groups, until there is no change in user-group assignments.

$PlayerA$ assigns sensed data points $(x_n)$ to user-group $(\mu_k)$ of $PlayerB$, such that the sum of the squares of distances of each data point to its closest user-group $(\mu_k)$ is minimum, which is expressed with the following objective function

$$J = \sum_{n=1}\sum_{k=1} r_{n,k}||x_n - \mu_k||^2. \qquad (7)$$

Therefore, Player $A$ uses an iterative procedure in which each iteration involves two successive steps(Lemma. 5 and Lemma. 6) corresponding to successive optimization with respect to $r_{nk}$ and $\mu_k$ for user-group initialization.

Lemma 5. *Player A (in stage 1), assigns n-th data point of observed data of IF(x) to the closest user-group of Player B*

$$r_{nk} = \begin{cases} 1, & if, k = argmin_j ||x_n - \mu_j||^2 \\ 0, & otherwise \end{cases} \qquad (8)$$

*Proof.* Since, $J$ is a linear function of $r_{nk}$, optimization yields a closed form solution. As terms involving different $n$ are independent, optimization is performed for each $n$ separately by choosing $r_{nk}$ to be 1 for whichever values of k gives the minimum value of

$$||x_n - \mu_k||^2 \qquad (9)$$

□

Lemma 6. *Player A (in stage 2), sets $\mu_k$ equal to the preference of all data points$(x_n)$ of observed IF(x) assigned to the user-group k.*

$$\mu_k = \frac{\sum_n r_{nk} x_n}{\sum_n r_{nk}} \qquad (10)$$

*Proof.* As objective function is a quadratic function of $\mu_k$, it can be minimized by setting its derivative with respect to $\mu_k$ to zero yielding

$$2\sum_{n=1} r_{nk}(x_n - \mu_k) = 0 \qquad (11)$$

which can be solved for $\mu_k$ to yield

$$\mu_k = \frac{\sum_n r_{nk} x_n}{\sum_n r_{nk}} \qquad (12)$$

where denominator is equal to the number of points assigned to usergroup $k$.

Let us validate learning model as an optimal solution of revenue-renewal problem (Theorem. 4), followed by the justification how *Whiz* quantifies the global performance of the likelihood estimation with EM (Theorem. 5).

Theorem 4. Learning model solves revenue-renewal problem having latent user-group as a mixture component of the Gaussian mixture model

*Proof.* Recruitment problem holds the key assumption that $COI(\theta)$ is distributed with multivariate Gaussian distribution, which lays out the technical foundation for devising *revenue-renewal* problem with Gaussian mixture model[53], in which, how the components are mixed in which proportions is not known a priori and hence, latent user-groups are represented by latent discrete variable. Hence, let us assume that a user-group is represented by a $k$-dimensional latent variable $z$, such that one particular $z_k$ is equal to one, where the rest of them are equal to zero. Hence, $z_k$ satiesfies the following

$$z_k \in \{0, 1\} \tag{13}$$

$$\Sigma_k z_k = 1 \tag{14}$$

Let us assume the distribution over this latent variable $z$ is multinomial, such as $P(z_k) = \pi_k$, where $\pi_k$ satiesfies the following

$$0 \leq \pi_k \leq 1 \tag{15}$$

$$\Sigma_{k=1}^{K} \pi_k = 1 \tag{16}$$

Therefore, given any $IF(x)$ and latent user-group ($z$), their joint distribution is defined in terms of their conditional distribution and the distribution of latent user-group ($z$) as follows

$$P(x, z) = P(z)P(x|z) \tag{17}$$

Hence, the marginal distribution of an $IF(x)$ is obtained by summing the joint distribution over all possible states of latent user-group ($z$).

$$P(x) = \sum_{z} P(z)P(x|z) \tag{18}$$

Since, $j$-dimensional $IF(x)$ is dependent upon $j$ $IF(x)$, each of which is measured from different latent $z$, the conditional distribution is expressed as

$$P(x|z) = \prod_{j=1} P(x_j|z) \tag{19}$$

However, this conditional distribution is written as follows

$$P(x|z) = \prod_{j=1}\prod_{k=1}^{K} N(x_j|\mu_{j,k}, \sigma^2_{j,k})^{(z^k)} \qquad (20)$$

Given user-group ($z$), only one term in product is active for all $k$, meaning $z^k$ is the selector, where
$z^k = 1$, for one index and $z^k = 0$, for others.
Hence, we obtain

$$N(x|\mu_k, \Sigma_k) = \prod_{j=1} N(x_j|\mu_{j,k}, \Sigma_{j,k}) \qquad (21)$$

Therefore, the marginal distribution over $IF(x)$ is defined as
$$P(x) = \Sigma_{k=1}^{K} \pi^k N(x|\mu_k, \Sigma_k) \qquad (22)$$

Hence, the marginal distribution of $IF(x)$ is formulated with Gaussian mixture having user-group($z$) as latent variable.

Finally, we obtain the estimable $COI(\theta)$

$$\theta = (\pi, \mu, \Sigma) \qquad (23)$$

$COI(\theta)$ is estimable parameter of EM-based unsupervised online Bayesian mechanism (in Section II).

Thus, learning model solves revenue-renewal problem having latent user-group as a mixture component of the Gaussian mixture model. □

**Theorem 5.** *Whiz quantifies the global bound of likelihood estimation achieved with original EM*

*Proof.* Whiz is equipped with a lower bound of the global MSE($MSE_G$) for COI($\theta$) estimation[50], which is random with a prior probability distribution. Hence,

$$MSE_G(\theta) = \int_{\Theta} MSE_L(\theta)p(\theta)d\theta \qquad (24)$$

□

Hence, *Whiz* quantifies the global bound of likelihood estimation achieved with original EM.

Let us justify *Whiz* as an optimal solution of *Online Stochastic fact-finding* (Theorem. 6), preceded by the validation how latest fact-finding research[42] is deterministic (Lemma. 13).

**Lemma 7.** *Fact-finding in[42] is a deterministic problem.*

*Proof.* Authors in[42] have devised CRLB[50] on MSE for fact-finding problem of estimating parameter (i.e. COI) ($\theta$) from the sensed data (x) (i.e. sensed data of the observed IF). However, this performs as a lower bound of the local MSE in true parameter value ($\theta_0$), such that $p(x|\theta_0)$ is the likelihood of the observed data(x) parameterized by true parameter($\theta_0$) and $\hat{\theta}(x)$ is an estimator of $\theta_0$. Since true value of parameter ($\theta_0$) in[42] is devised as deterministic (Definition. 5), fact-finding in[42] is a deterministic problem. □

Theorem 6. *Whiz is an optimal solution of Online Stochastic Fact-Finding*

*Proof.* In comparison to local MSE ($MSE_L$)[50], which is achieved with deterministic fact-finding (Lemma. 13), Whiz is equipped with a lower bound of the global MSE ($MSE_G$)[50] for $COI(\theta)$ estimation (Theorem. 5), where $COI(\theta)$ is random with a prior probability distribution. Hence, Whiz is an optimal solution of Online Stochastic Fact-Finding. □

## 4. Empirical Results

Extensive experiments are performed on synthetic datasets to justify the supremacy of *Whiz*. Its estimation accuracy and convergence are justified over deterministic counterparts of *revenue-renewal* (i.e. Tube[45]) and *fact-finding* (i.e. EM[41], Bayesian[43], Voting[46], Average-log[54], Sums[54], TruthFinder[59] and PageRank[47]) schemes. Then, its estimation accuracy is evaluated by comparing its false positives and false negatives with true positives and true negatives, respectively.

### 4.1. Comparison with counterpart revenue-renewal model(i.e. Tube)

The supremacy of *Whiz* (denoted as *Whiz*) is evaluated over a counterpart *revenue-renewal* model, namely Tube (denoted as Tube[45]) in terms of the estimation accuracy.

#### 4.1.1. Experiment set-up

Tube[45] monitors Smart-device usage between time-independent and time-dependent schemes to estimate optimal waiting function parameters, so Smart-device users defer an application session for a given amount of time ($\tau$) for a given discount ($d$) from some baseline metered price. However, the waiting function is parameterized to quantify the various price-delay trade-offs corresponding to different users and application sessions. Moreover, this waiting function is decreasing in time deferred ($\tau$) and concave and increasing in the discount offered ($d$), since users are less likely to defer as time deferred increases but are more likely to defer if offered a larger monetary reward

$$w_p(d, \tau) = \frac{d}{\lambda_p(\tau + 1)^\rho} \quad (25)$$

where $\rho$ is a parameter measuring patience, the patience index and $\lambda_p$ is an appropriate normalized constant.

Therefore, we have devised Tube[45] as a counterpart deterministic *revenue-renewal* model with EM having observation and parameter summarized in Table. 2. However, *Whiz* is devised with EM with BCRLB to estimate *COI* from the Smart-device data of observed *IF*. In our experiment, the actual estimation variance of *COI* is obtained from the average root mean square error (RMSE) over all *IF*.

| Terminologies | Notation(Tube) | Notation(Whiz) |
|---|---|---|
| Observation | Discount ($d$), Defer-time ($\tau$) | IF ($x$) |
| Parameter | Patience ($\rho$) | COI ($\theta$) |

Table 2: Terminologies and Notations in Tube[45] and Whiz

*4.1.2. Results*

In this first experiment (Fig. 4(a)), the estimation accuracy of *Whiz* is compared with that of Tube[45] by varying the number of *IF*. The results are averaged over 100 experiments. The RMSE remains much smaller than the Tube, even when the number of *IF* increases. The RMSE remains much smaller than Tube throughout the time. However, as the number of *IF* increases, *Whiz* gradually tracks the variation of *COI* more accurately and converges to the RMSE. Hence, The higher number of *IF* leads to the higher availability of the observed data for more accurate quantification of *COI* estimation in *Whiz*.

In the second experiment (Fig. 4(b)), the estimation accuracy of *Whiz* is compared with that of Tube[45] by varying the number of *COI*. Tube seems to be higher than the RMSE throughout. However, as the number of *COI* increases, *Whiz* tracks the RMSE more accurately and soon converges to it. Since the higher number of *COI* helps in acquiring sufficient Bayesian knowledge, more accurate quantification of *COI* estimation is achieved in *Whiz*.

In the third experiment (Fig. 4(c)), the estimation accuracy of *Whiz* is compared with that of Tube[45] by varying the satisfaction ratio of *IF*. The satisfaction ratio of *IF* denotes its probability of successfully estimating *COI*. However, this ratio is varied from 0 to 1 in our experiment. Then, the results are averaged over 100 experiments. As the satisfaction ratio of *IF* increases, both schemes gradually tend to converge to the RMSE. However, *Whiz* tracks the estimation variation more accurately than Tube. Hence, the higher satisfaction on *IF* leads to better quantification of *COI* estimation, which is, in fact, achieved through the Bayesian offline filtering mode of *Whiz*.

In the fourth experiment (Fig. 4(d)), *Whiz* is compared with Tube[45] by varying the demand frequency of *IF*. The demand frequency of *IF* denotes the frequency of observations it makes. Hence, a 10 observations is regarded as the demand ratio of 1 during the experiment. Then, the results are averaged over 100 experiments. As the demand ratio increases, both schemes tend to converge to the RMSE. Hence, frequent observations lead to the better quantification of *COI* estimation, which is achieved through the Bayesian online filtering mode of *Whiz*.

*4.2. Comparison with state-of-the-art deterministic fact-finders*

The supremacy of *Stochastic Monetization* property of *Whiz* is compared with the state-of-the-art deterministic fact-finders, namely Voting[46], Page-rank[47], Average-Log, Sums[54], Truth-finder[59], Bayesian fact-finder[43] and EM-based fact-finder[41] in terms of the estimation accuracy.

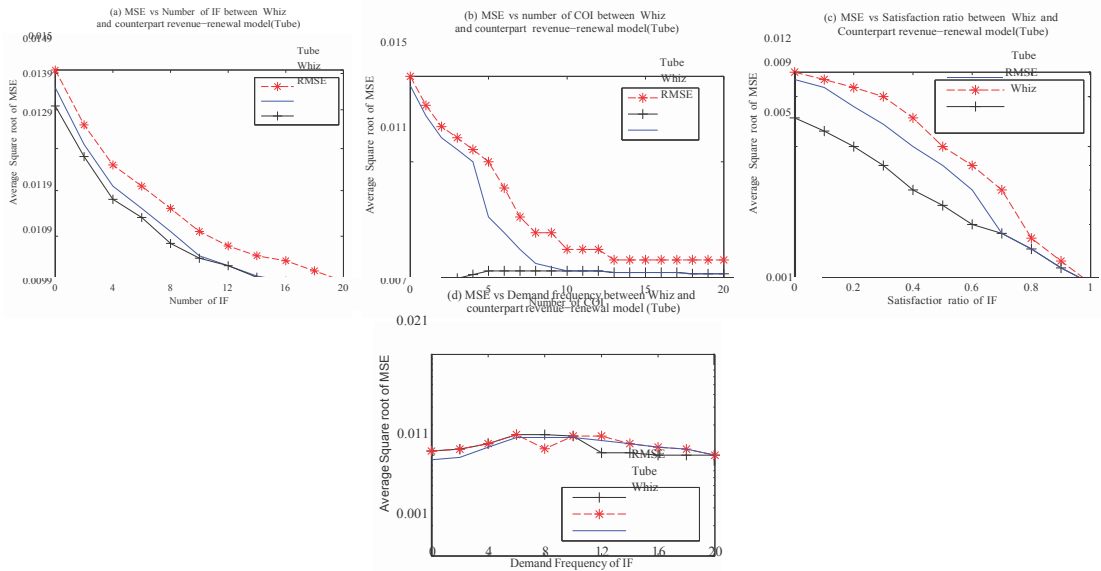

Figure 4: Comparison of estimation accurary between Whiz and a counterpart revenue-renewal model (i.e Tube[45])

### 4.2.1. Experiment Settings

A simulator is set on Matlab to generate a random number of IF and COI. The satisfaction-ratio of an IF is represented by a probability $P_1$, which is assumed to be randomly generated. On the other hand, the satisfaction-ratio of a COI is represented by $P_2$, which is assumed to be uniformly distributed between 0.5 to 1.

In the experiment, the estimation accuracy of IF satisfaction is obtained by comparing the computed ground truth probability with the ground truth probability, that holds the truth. However, the estimation accuracy of COI satisfaction is obtained from two metrics, namely false positives and false negatives. The false positive is denoted by the ratio of number of false COI, that are classified as true, over total number of COI, that are classified as true. However, the false negative is denoted by the ratio of true COI, that are classified as false, over the total number of COI, that are classified as false.

### 4.2.2. Results

In the first experiment (Fig.5(a)-5(c)), the estimation accuracy of Whiz is compared with that of state-of-art fact-finders by varying the number of IF. Hence, the number of COI is kept fixed at 50, of which 25 are satisfactory and 25 are unsatisfactory. However, the number of IF is varied from 1 to 10. Moreover, each IF is assumed to be related with 10 COI. Then, the results are averaged over 100 experiments. As the number of IF increases, Whiz provides much lower estimation error than the state-of-art fact-finders. Since the inclusion

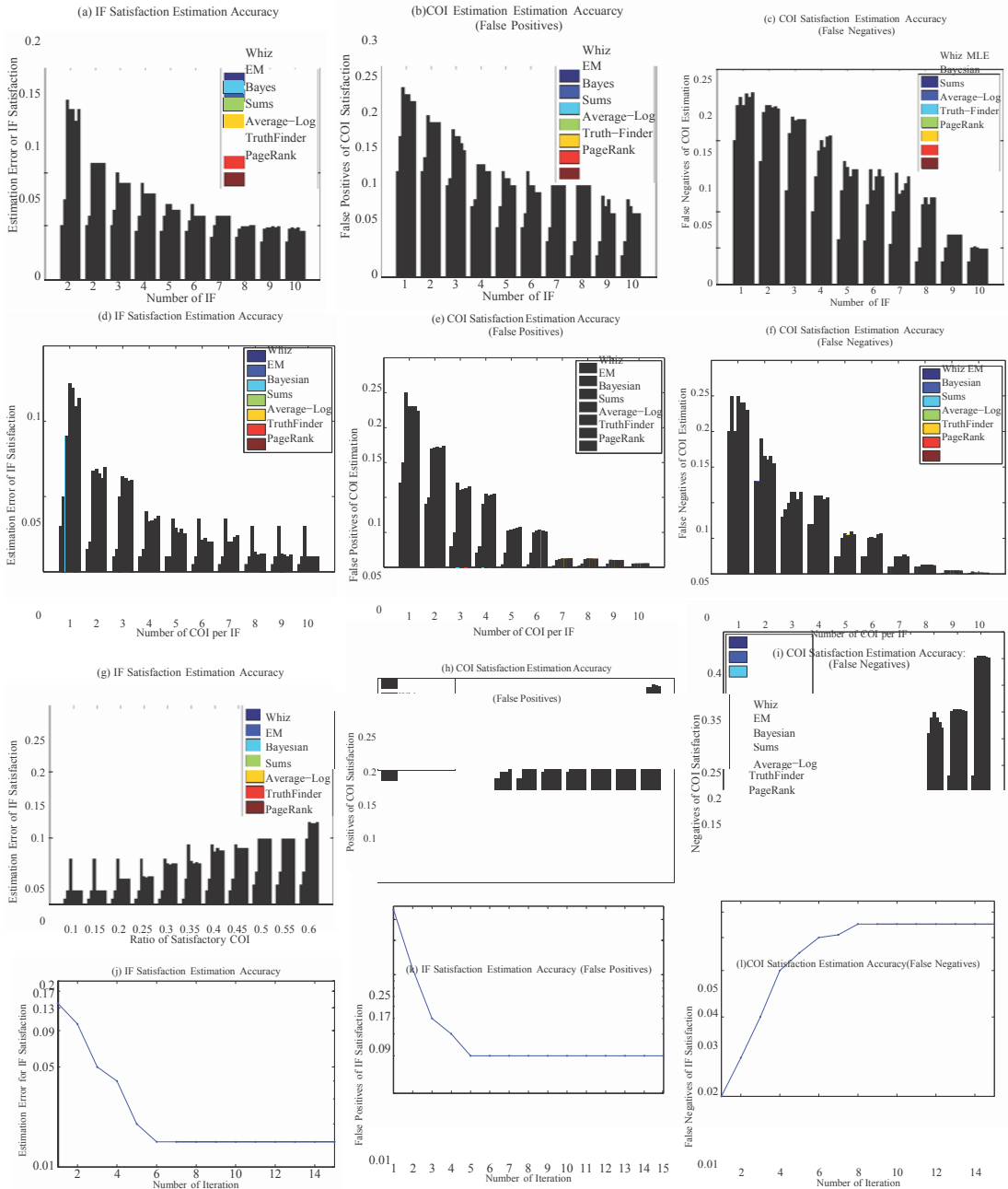

Figure 5: ((a)-(i)) Comparison of estimation accuracy between *Whiz* and counterpart monetization schemes baseline fact-finders, and ((j)-(l)) convergence of *Whiz*

of *IF* enhances the chance of the availability of the Bayesian knowledge about *COI*, *Whiz* achieves better estimation accuracy than others.

In the second experiment (Fig.5(d)-5(f)), the estimation accuracy of the *Whiz* is compared with that of other fact-finders, when the number of *COI* per *IF* changes. Hence, the total number of satisfactory and unsatisfactory *COI* are kept 25 respectively. Moreover, the average number of *COI* per *IF* is kept 1 to 10, by assuming there are total 50 *IF*. Then, the results are averaged over 100 experiments. The higher number of *COI* per *IF* leads to the gradual degradation of the estimation accuracy in all schemes. However, *Whiz* uses Bayesian knowledge to distinguish the satisfactory *COI* from the unsatisfactory one and thus globally quantifies the likelihood to achieve higher estimation accuracy than others.

In the third experiment (Fig.5(g)-5(i)), the impact of the ratio of the satisfactory *COI* is compared among all schemes. Hence, the ratio of the satisfactory *COI* is varied from 0.1 to 0.6, while keeping the total number of *COI* fixed at 50. Moreover, the number of *IF* is kept fixed at 50, while keeping the average number of *COI* per *IF* 10. Then, the results are averaged over 100 experiments. *Whiz* shows almost the similar performance like other schemes, when the ratio of the satisfactory *COI* is small. However, as this ratio increases, *Whiz* gradually achieves higher estimation accuracy. Since this higher satisfactory ratio helps in acquiring better Bayesian knowledge about *COI*, the global quantification of the likelihood estimation is achieved with *Whiz*.

In the fourth experiment (Fig.5(j)-5(l)), the convergence property of *Whiz* is evaluated in terms of the estimation error of *IF* satisfaction and *COI* satisfaction. Hence, the number of both satisfactory and unsatisfactory *COI* is kept 25. Moreover, the number of *IF* is kept 50, while keeping the average number of *COI* per *IF* 10. Then, the results are averaged over 100 experiments. Initially the presence of unsatisfactory *COI* leads to higher estimation error. However, as soon as Bayesian knowledge is achieved about *COI*, *Whiz* achieves better estimation accuracy and converges. Hence, as the iteration increases, *Whiz* converges with Bayesian global quantification of the likelihood estimation of *COI*.

### 4.3. Estimation accuracy from False Negatives/False Positives

The estimation accuracy of *Whiz* is evaluated by comparing its false positives and false negatives with true positives and true negatives, respectively, which are computed from the ground truth.

### 4.3.1. Results

In the first experiment (Fig. 6(a) and Fig. 6(e)), the estimation accuracy is evaluated through false positives/negatives by varying the number of *IF*. Hence, the number of true and false *COI* are kept fixed at 1000, while keeping the average number of observations per *IF* 300. However, the number of *IF* are varied from 0 to 20. Then, the results are averaged over 1000 experiments. As the number of *IF* increases, false positives and false negatives track the actual values more accurately and gradually decrease to the minimal value. Hence, the higher number of *IF* leads to the better chance of the availability of the observed knowledge, which is required for the quantification of the *COI* estimation.

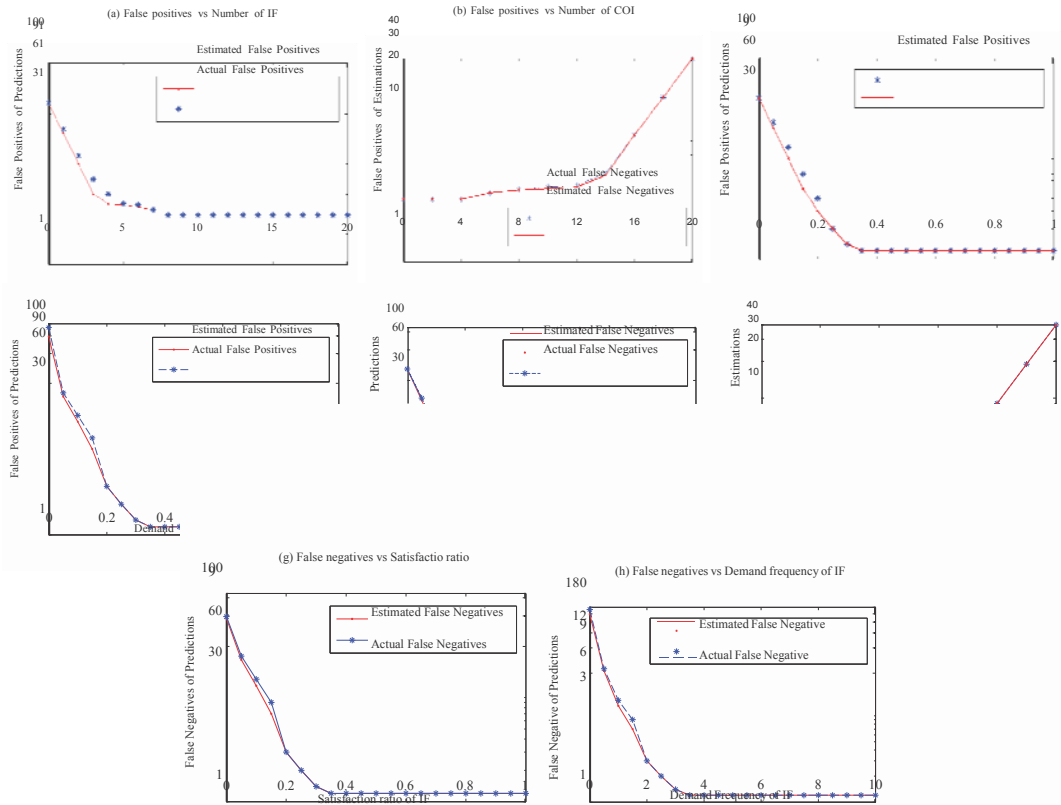

Figure 6: Estimation accuracy of $Whiz$ in terms of False Negatives and False Positives

In the second experiment (Fig. 6(b) and Fig. 6(f)), the estimation accuracy is evaluated through false positives/negatives by varying the number of *COI*. Hence, the number of *IF* is kept fixed at 30, while keeping the average number of observations per *IF* 100. However, true and false observations are kept the same. Moreover, the number of *COI* is varied from 0 to 20. Then, the results are averaged over 1000 experiments. As the number of *COI* increases, false positives and false negatives track the actual values more accurately, however, their estimation performance degrades. Since the number of *IF* and observation per *IF* remain the same, the increasing number of *COI* leads to the degraded performance in estimation.

In the third experiment (Fig. 6(c) and Fig. 6(g)), the estimation accuracy is evaluated through false positives/negatives by varying the satisfaction ratio of *IF*. Hence, the number of *IF* is kept 30, while keeping the number of true and false observations 1000. However, the observation per *IF* is set to 100. Then, the satisfaction ratio of *IF* is varied from 0 to 1. As the satisfaction ratio of *IF* increases, estimated false positives and false negatives track the actual values more accurately and decrease to converge. Hence, the higher satisfaction ratio of *IF* leads to the higher chance of the accurate quantification of *COI* estimation.

In the fourth experiment (Fig. 6(d) and Fig. 6(h)), the estimation accuracy is evaluated through false positives and false negatives by varying the demand ratio of *IF*. Hence, the number of *IF* is kept at 50, while keeping the number of true and false *COI* 1000. However, the demand ratio is varied from 0.1 to 1. Then, the results are averaged over 1000 experiments. As the demand ratio of *IF* increases, estimated false negatives and false positives track the actual values more accurately and decrease to converge. Frequent observations made by *IF* facilitate more Bayesian knowledge about *COI*. Hence, the higher demand ratio of *IF* leads to the accurate quantification of the *COI* estimation.

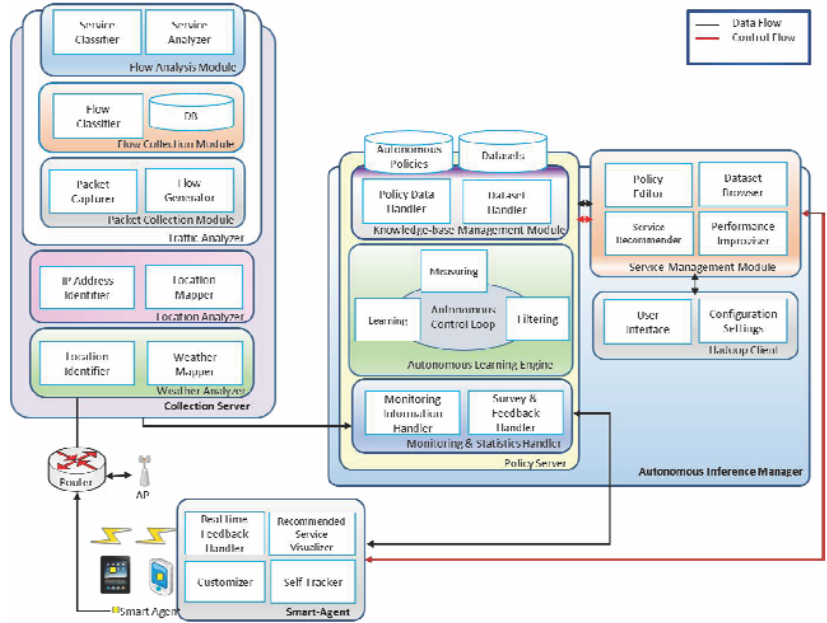

Figure 7: System Architecture

*4.4. Personalized IoT-Service Recommendation: Data Collection, Prototype Development and YouTube-usage dynamics Approximation*

A survey is undergone among 78 academicians (i.e. 39 students, 22 alumni and 17 school officials) about their Smart-device usage-pattern in different contexts, such as emotion, weather, location and time of a day[60]. It is followed by prototype app installation on their Smart-devices to monitor traffic pattern and to incorporate their real time feedback in different weather, location and time. However, as it is difficult to infer emotion from traffic-patterns, E-mail survey dataset[60] results are utilized to infer their behavioral pattern. Hence, this offline dataset is utilized for ground truth selection in prototype development for personalized service recommendation in an example *Happy IoT* scenario[60]. Then, dynamic usage-contexts among YouTube video user-groups are approximated with *Whiz* on YouTube dataset[55]. Finally, filtering of rumor spreading trend among dynamic user-groups in social network is approximated with *Whiz* on Higgs Twitter dataset[56].Moreover, approximation on datasets are performed with Psychic-inference-engine[60], a Python-based library for multivariate Gaussian distribution. Python is used to utilize the benefit of open source scientific computational-tools like Numpy, Scipy, Matplotlib, etc, since these libraries support computations involving matrix operations, like variance-covariance, inverse and transpose.

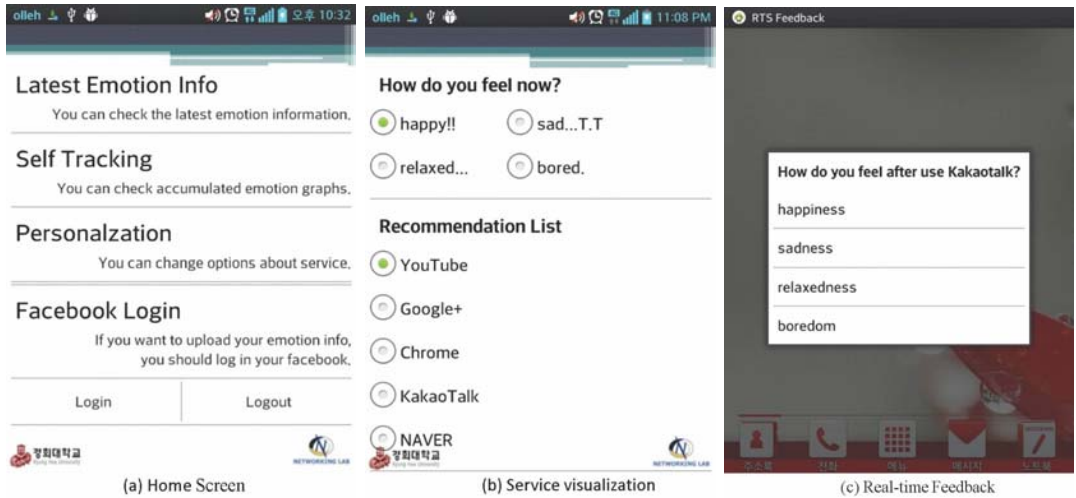

Figure 8: Smart Agent

### 4.4.1. Data Collection

Primary survey questionnaires are intended to discover participants passion about various Smart-devices, such as Smart-phone (94.1%), Smart-tab (41.2%), Smart-TV (11.8%) or even Smart home-appliances(5.9%)[60]. Most of them prefer Smart-devices as an essential companion (64.7%) or entertainment media (58.8%). A glimpse of their entertainment affection is that most of them always(35.3%) or sometimes (47.1%) enjoy radio/ TV-programs. Among them, the majority of them prefer such entertainment prior to sleep (47.1%), however, the rest enjoy entertainment at either travelling (from/to workplace) (41.1%) or dinner time (29.4%). Not surprisingly, their affection to entertainment (58.8%), news (58.8%) or drama (29.4%) and even sports (35.3%) is an indication of priority on multimedia Big Data in Smart-device traffic. Participants are asked to guess about their possible emotion (e.g. happiness, sadness, boredom, relaxation) in different parts of a day. Moreover, they are instructed to predict preferred services (e.g. Music, Chatting, Browsing, Social Networking, TV/Radio program, Gaming) in their different emotions. Questionnaires are also intended to retrospect the discrepancy of emotions and relevant services in different weather conditions, such as sunny, snowy, rainy and cloudy. Participants are asked about their preferred services at travelling (e.g travelling to or from workplace) or stationary (e.g at workplace or home) moments. Consequently, location-based usage-pattern is apparently an indication of usage-preferences in different parts (e.g. morning, afternoon, evening, night) in a day. Moreover, questionnaires are set to derive behavioral service usage-pattern specially in social network-based services, which have recently emerged as popular items. In this context, questionnaires resemble participants interest about music (47.1%), comedy (29.3%) or even drama (35.3%) in social gathering, which more or less indicate consumer-affection for entertainment in social network. Consequently, a large portion (53.3%) are reported to engage themselves in sharing music/video/photo in social network. Moreover, the majority (64.7%) express that they generally share their emotions through social networks.

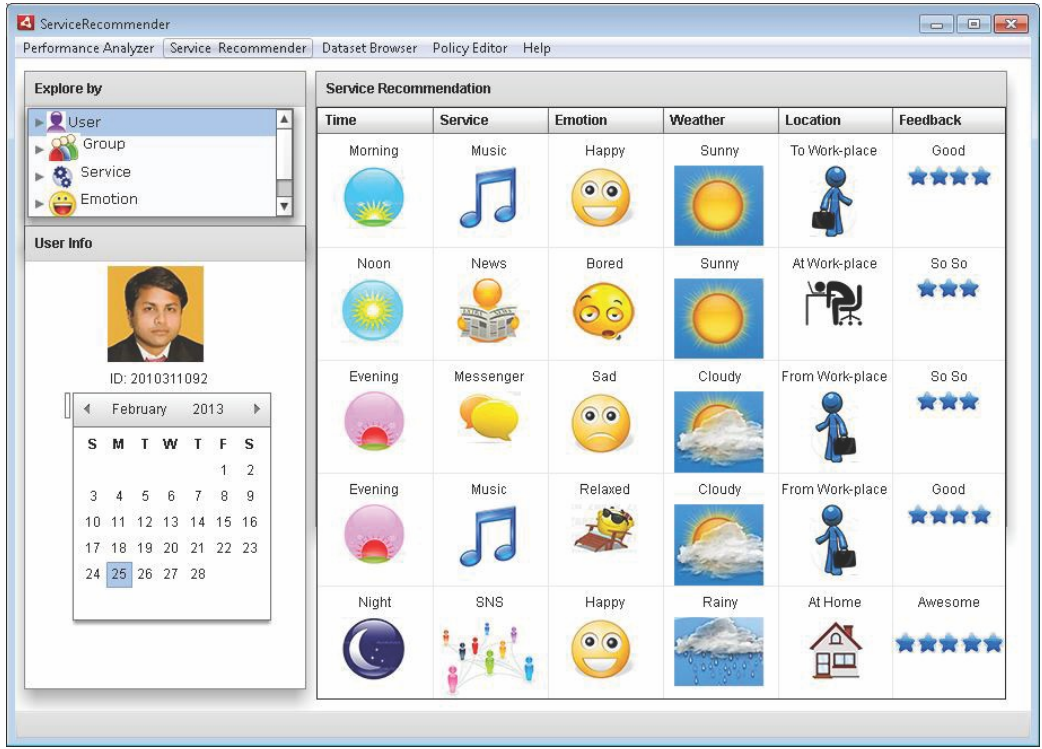

Figure 9: Web UI prototype in service recommendation scenario for work-going persons.

*4.4.2. Prototype Development*

Prototype is developed to infer Smart-device users usage-dynamics in different weather, time and location; recommend service and meanwhile, incorporate their real-time feedback. Prototype includes Smart-Agent, Collection Server, Autonomic Inference Manager and Hadoop-based Traffic Measurement and Analysis Platform (Fig. 7)[60]. Smart-Agent (Fig. 8), an Android client installed on Smart-device, tracks users time and environmental information. However, whenever users visualize recommendation on their interest (i.e. COI), they either execute or decline recommended service and consequently give real-time feedback. Based on earned satisfaction on personal interest, their feedback is awesome, good, so so or even worse. Hence, Smart-device users service-execution, declinement, usage-time, usage-duration and real-time feedback are functional as IF for service providers. Collection server is equipped with three major sub-modules, namely Traffic Analyzer, Location Ana-lyzer and Weather Analyzer. Traffic Analyzer is developed with Jpcap to analyze network traffic of Smart-device users. However, Location Analyzer is developed with KISA location API to extract the location information through IP address. Moreover, Weather Analyzer is developed with Yahoo weather API to extract the weather information. On the other hand, Autonomic Inference Manager is equipped with two major sub-modules, namely Policy Server and Service Management Module. Policy Server incorporates autonomous learning engine to learn service-recommendation, manage policy and datasets for a service management system. Service Management Module, on the other hand, is used by service-providers to visualize user-statistics, edit/configure/view service recommendation policies, improvise system performance, etc. Hadoop-based traffic-measurement and analysis platform maintains datasets of traffic-analysis history and offline survey. As it is very difficult to predict user-emotion from traffic-analysis, survey questionnaires are dedicated to know about peoples preferred services in different emotions for different location or weather on even different parts of a day.

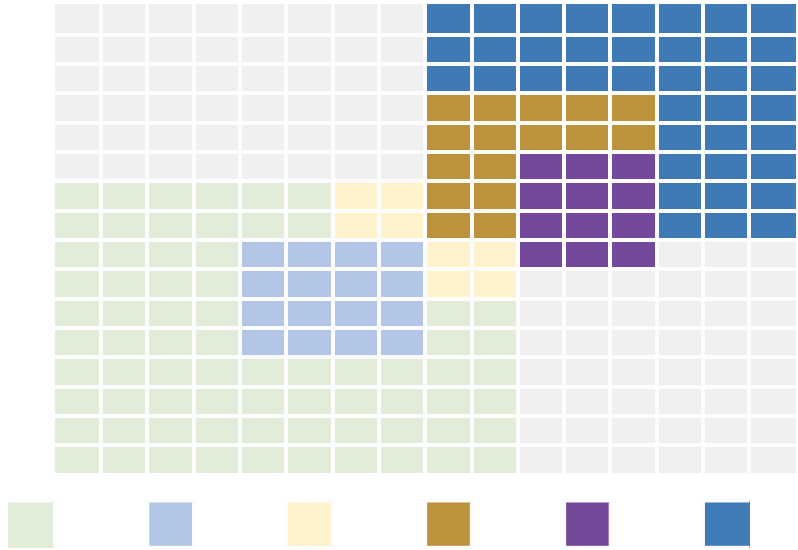

Figure 10: Co-occurrence matrix for inferring different video user-groups in YouTube dataset

*4.4.3. YouTube Video user-groups approximation*

YouTube dataset[55] is trained with Whiz to approximate user-groups (i.e. COI) for different video categories (Fig. 10). Hence, latent 5 user-groups, namely music, drama, sports, travel are considered as COI in experiment. Even though YouTube-usage dynamics is inferred from number of views, shares, total watch time, even number of subscribers of the channel or even duration of statistical analysis, total number of views is assumed as IF in our experiment. Music is preferred, when participants are either travelling from/to workplace. Among them, majority enjoy reality-shows, when they are relaxing at home. However, the rest, more or less, enjoy comedy shows. On the other hand, drama is preferred, when they are taking rest at home or travelling from/to work. Among them majority prefer browsing different channels, when they get lunch-break. However, the rest prefer news, when they are travelling from/to workplace.

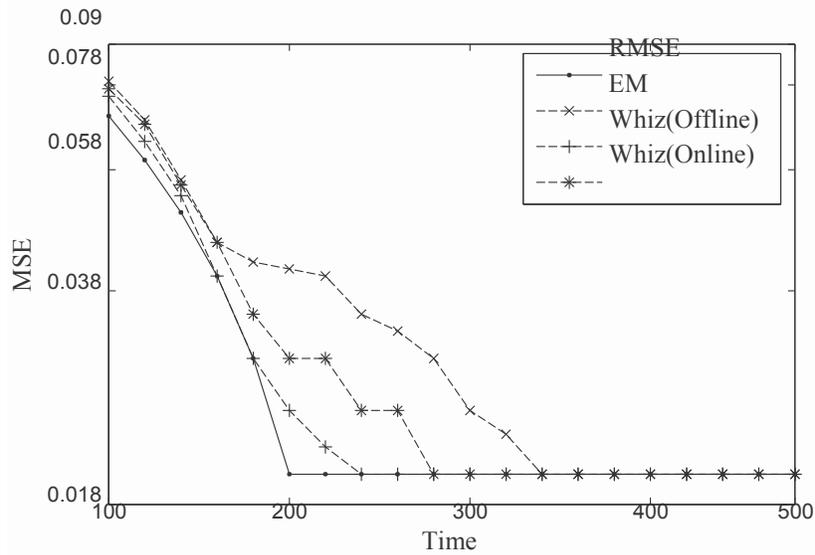

Figure 11: Estimation accuracy with accelerometer traces

*4.4.4. User Activity Approximation*

Cenceme[61] dataset is trained with *Whiz* to approximate user activity from the observed uploading data of users, so that smart-phone users upload real time sensed data to server, by maintaining trade-off between accuracy/battery consumption. Cenceme dataset contains 2 weeks' sensed information (i.e. accelerometer raw data and GPS location coordinates) of 20 Nokia N95 users (i.e. students and staff members) of the department of Computer Science and Biology at the Dartmouth college. Users activities (i.e. COI)(e.g sitting, running, walking, standing) are estimated by using Whiz from observed data of Accelerometer and GPS traces. Accelerometer raw traces are observed from Timestamp (i.e. the time when the line has been written into the log file) and $X_{acc}$, $Y_{acc}$, $Z_{acc}$, the accelerations on the three axes. It is observed (Fig. 11) that Whiz converges to RMSE pretty faster than conventional EM with accelerometer data. However its offline version gives better estimation accuracy, since continuous movement/motion often causes deviation from accuracy. On the other hand (Fig. 12), Whiz converges to RMSE relatively slower than EM with GPS data (e.g. altitude,latitude,longitude,hdop,speed*altitude,latitude,longitude ,hdop,speed). However, it online version yields better estimation accuracy, since GPS are difficult to process, as often there might be no samples. Moreover, offline version works better in situation like user is sitting for long period (e.g. 15 minutes) and users last known position is significant in this quantification.

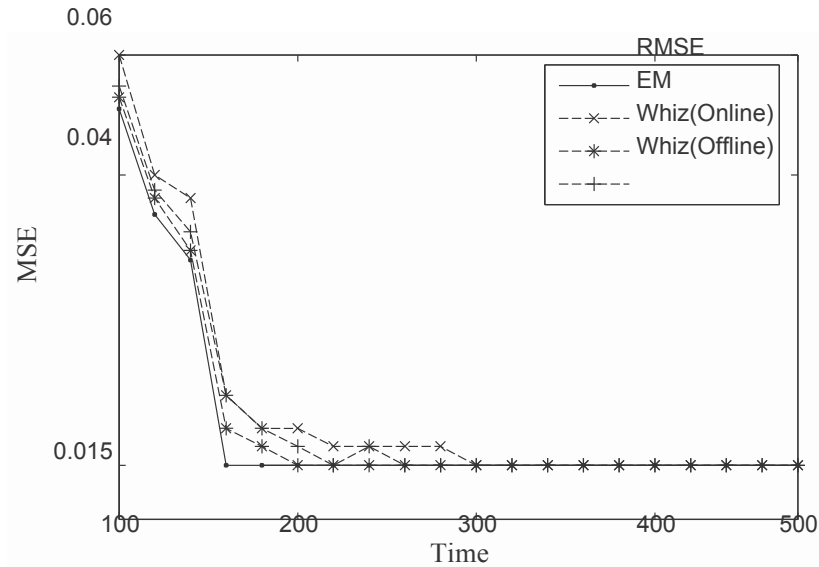

Figure 12: Estimation accuracy with GPS traces

*4.4.5. Social Rumor Trend Approximation*

The Higgs Twitter dataset[56] is trained with Whiz to approximate the trend of Social Rumor spreading at a scientific event (Fig. 13). The Higgs dataset includes Twitter messages before, during and after the announcement of discovery of a new particle on 4th July, 2012. The dataset consists of 985590 Tweets collected between 1st July 2012 to 7th July 2012, which include one of the following keywords, such as ihc, cern, boson and higgs. The resulting social network graph consists of 456631 sources (i.e.Tweet authors) and 14855875 directed edges (i.e. follower/followee relationship). Accuracy of rumor estimation is filtered among four latent user-groups, who are categorized based on time intervals on rumor-spreading. It is observed that (Fig. 11), higher estimation accuracy is achieved among the first user-group, since rumor spreads faster at beginning. Then, the lowest estimation accuracy is achieved after the pre-announcement, since rumor spreads at the slowest rate. However, the estimation accuracy is increased a bit gradually, since there are still curiosity among people before final announcement. At long last, relatively higher estimation accuracy is achieved, since there is a common trend of diffusion of rumor across the other end of the world as time elapses.

In this first experiment, the estimation accuracy is evaluated by varying the number of observed IF (i.e. Retweeting, Replying, Mentioning, Following). The results are averaged over 50 experiments. It is observed that offline converges faster than online to RMSE. Since online version needs to continuously regulate Tweeter sources to acquire better knowledge about rumor trend, they converge to RMSE gradually. However, offline version, after few intervals, regulate knowledge about rumor event from the observations on retweeting, replying and mentioning neighboring nodes and therefore, converge faster.

In the second experiment, the estimation accuracy is evaluated by varying the number of COI (i.e. social user-groups on different time periods, such as before pre-event, during pre-event, post-pre-event and before main-event, during-main-event and after-main-event ).The results are averaged over 50 experiments. Even though both online and offline versions converge to RMSE, offline does it in a faster way. However, as COI increases, offline quantification assists capturing similarity in rumor trends, especially between two pairs of user-groups (pre-pre-event and after-pre-event) and (pre-main event, after-main-event).

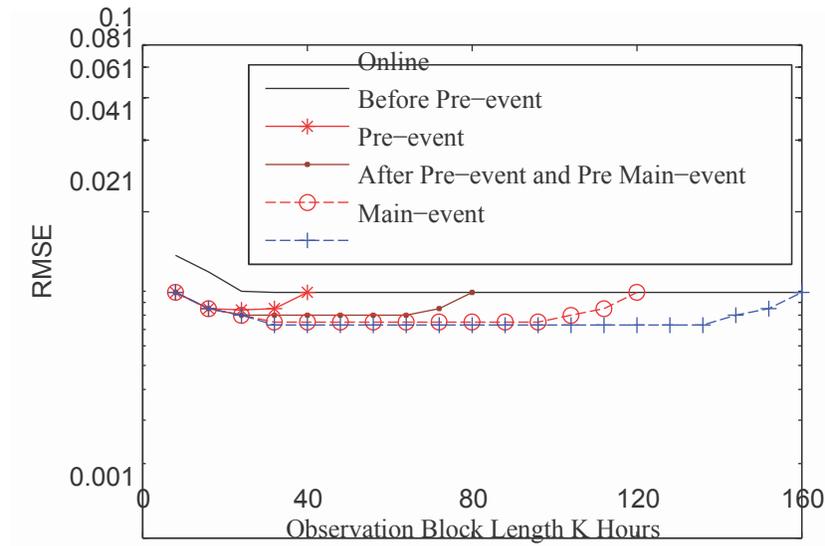

Figure 13: Filtering social rumor trend with Online and different Offline modes of Whiz

In the third experiment, the estimation accuracy is evaluated by varying the satisfaction ratio of observed data source(i.e. Retweeting, Replying, Mentioning, Following). The satisfaction ratio denotes its probability of successfully estimating COI. However, this is varied from 0 to 1 in our experiment. Then, the results are averaged over 50 experiments. Eventually, as this ratio increases, both schemes gradually converge to RMSE, however online version seems to be faster. The increase in satisfaction of observed IF assists in gradually acquiring better knowledge in COI estimation.

In the fourth experiment, the estimation accuracy is evaluated by varying the frequency of the observation an IF performs. In this context, a 100 observations is regarded as the demand ratio of 1 during the experiment. Then, the results are averaged over 50 experiments. As the demand ratio increases, all schemes gradually converge to the RMSE. However, online converges relatively faster than offline. The frequent the Tweets are observed in between first pre event and main event, for example, the better is the chance of acquiring sufficient knowledge about main event.

## 5. Related Works

*Autonomic Happy IoT Management* necessitates service providers to synchronize to the dynamic usage-contexts of subscribers. In this context, service providers design innovative incentivization mechanisms to motivate subscribers and thereby measure the immediate response through some incentive filters. Hence, unsupervised online Bayesian mechanism is turned out to be a solution of stochastic fact-finding of iteratively estimating fact(usage-context) from the source(incentive-filter), when fact is assumed to be random with a prior probability distribution. In this context, state-of-art of IoT-driven context-aware services, innovative incentivization mechanisms and incentive filters; fact-finding and Bayesian synchronizers are presented.

### 5.1. Context-awareness in IoT Services

Usage-contexts(emotion, weather, location, etc), inferred from ambient sensed information of subscribers smart-devices, are frequently utilized by IoT service providers for context-aware services.

#### 5.1.1. Emotion-awareness

Emotional mode inferred from smart-device traces are used in personal search[3], well-being[4] or personalized shopping applications[5]. Behavioral pattern of individuals and groups is often inferred from mobility-pattern, weather, time and activity to facilitate a personalized search experience for smart-device users[3]. Moreover, on/off-campus mobility pattern, collected from Smartphone sensors, are utilized to infer individual behavior pattern for well-being care at sickness[4]. Global behavioral pattern(idle walking, fast walking and stopping) of smart-phone users in shopping malls are used to predict the near future behavior for a personalized shopping experience[5]. Moreover, psychological states of Smartphone users play a vital role, if an obtrusive way is applied, for example, e-mail survey on privacy concern about Smartphone apps[62]. Gaussian mixture models, collaborative filtering are recognized as prominent tools to infer emotions in IoT-enabled services. For example, Gaussian mixture model is used to infer the correlation between activity and emotion and emotion and location among individuals and groups from smart-phone usage-history[39]. Moreover, evolvable classification and collaborative inference are devised to predict the human behavior from the smart-phone usage-contexts[63].

#### 5.1.2. Location-awareness

Location traces from the smart-devices are used in traffic signal detection [6][7], transit tracking[8][9], localization[10][11], sociability detection[12], consumer-care[13][14][15], media-sharing[16] and transportation[17]. In this context, supervised and unsupervised classifiers[6], support vector regression model and low-pass and collaborative filters[7] are frequently used to detect traffic signal from Smartphone GPS traces. However, online filtering[8], activity classification, spatio-temporal route-matching and hidden markov model[9] are applied in transit tracking from Smartphone GPS traces. Support vector machine, color clustering-based classification[10], topic models[11], reinforcement learning[12] are used for identifying logical location, characterizing places and recognizing sociability in office environment from ambient sensor reading of smart-devices. GPS localization scheme is also devised in to recommend the cheaper grocery product and store location Smartphone users[13]. Moreover, markov decision model[14], hidden markov decision model and adaptive sampling schemes[15] are devised to infer the mobility pattern of Smart-device WiFi tracks to find missing mobile devices and to predict the optimal user-experience in wireless connectivity. Often estimation theory[16], collaborative filtering[17] are used for inferring colocation patterns in media content-usage among fellow commuters and predicting bus route or arrival time from the surrounding environmental contexts of Smartphone users.

*5.1.3. Environment-awareness*

Smart-device traces are used in environmental report generation[18][19][20], event detection [21][22][23]. Regression model[18], hidden markov model[19], compressive sensing[20] are frequently used to measure air pollution, generate environmental impact report or noise map by quantifying the relation between environmental characteristics(local traffic, population and weather) and smart-device traces. However, online decentralized anomaly detection[21], Bayesian statistics[23] are also devised for smart-device based earthquake and nuclear threat monitoring in a city. Moreover, resource metering, forced amnesia, sensor taint tracking and access control mechanisms[22] are devised for smart-device based party thermometer application.

*5.2. Incentivization*

Innovative incentive mechanisms are designed by IoT-service providers to motivate subscribers with monetary support [24][25][26], social reputation [27][28][29], quality-of-information assurance[30][31] or even enhanced consumer care [32][33][34]. Stackelberg game[24], reverse-auction[25], micro-payment-based[26] incentivization schemes reserve the lowest service charge for subscribers, motivate altruism and even encourage competitiveness. Repeated game, simple heuristic based incentive mechanisms incentivize subscribers for complying with the prescribed social strategy[27], maintaining individual fairness and social welfare[28] and thereby sometimes reward comes as increased reputation in the social network[29]. Optimal auction[30], game theory[31] benefice those subscribers, who have given useful and accurate information in recent time. Simple heuristics[32], greedy[33] incentivization schemes often motivate subscribers through context specific 3D-budget utilization and even mobility aware search experience. Even psychological insights[34], such as human demands are often flexible to genre, accessibility and price of incentive; often fewer choices simplify decision making of participant are utilized by service providers to incentivize subscribers, for example, by prefetching mode-specific media content in a Wi-Fi hotspot.

*5.3. Incentive Filter*

Conventional incentive-filter/popularity-measurement-tools (download-count, user-rating, usage-duration) are now renovated [35][36][37] to adapt to the dynamic IoT service-usage. Prepp[35] is such an effort, which predicts not only which Smart-device app might be used, but also when it might be used. Appjoy[36] does it with a collaborative filtering technique. However, Pulse[37] translates qualitative user-reaction(frequent head-movement/talking or lack of attention during movie) into quantitative user-rating. The increasing demand of internet media traffic also introduce video-engagement-measurement tools[64], such as startup latency, genre of buffering event, the quality of video. Moreover, ad-supported and subscription based media-traffic business models necessitate additional popularity measurement tools, such as ad-playing and user-retention time, etc. However, user-interaction, application-usage, network traffic and energy are often utilized together to derive not only quantitative variance, but also the qualitative similarities among usage-behavior. In this context, dynamic usage-behavior is often monitored through the network edges, which are scalable, flexible and free of the burdens of centralized control[65].

*5.4. Fact-Finding*

Even though the early age of fact-finding research is characterized by the weighted ranking, recent relevant efforts are renovated by the inclusion of credibility with the advent of statistical learning. Voting[46], an earlier fact-finder, is believing on the fact reported by the majority sources. Hence, the most voted facts are often believed over relatively few reliable facts. Similarly, PageRank[47] rates a web-page as popular by calculating its many backlinks or a few highly ranked backlinks. Hence, not-so-popular web-pages are often more accurate than the popular one. In this context, fact-finding[54], characterized by nodes(sources, claims), edges(who claims what), is introduced as an iterative weighted ranking process for sources and claims. However, the credibility of neither the source nor the claim is achievable through such weighted-sum approach. In this context, TruthFinder[59] is introduced to iteratively infer the trustworthiness of websites and facts from the simple relationships between different claims. Bayesian interpretation quantifies the probability that a source is truthful or a claim is true in the absence of the detailed prior knowledge[43]. However, this scheme is very sensitive to initial conditions of iterations. Maximum likelihood estimation calculates the reliability of fact without any prior knowledge of the reliability of sources[41]. However, the desired estimation performance is not guaranteed with such local likelihood estimation. In this context, Cramer-Rao lower bound quantifies the error of an unbiased fact estimator in the deterministic setting[42] . However, the scheme is based on the assumption of independent facts and the hardness of fact-estimation is not taken into consideration. However, the global quantification of the likelihood estimation is not achievable with such deterministic lower bound. Attempts are undergoing to incorporate prior knowledge into basic fact-finder, through several extended algorithms, namely average-log, investment, pooled investment. The notions of hardness of claims is also introduced into fact-finding through algorithms, namely cosine, 2-estimates and 3-estimate[66].

*5.5. Bayesian Synchronizers*

Lower bound based suboptimal synchronizers [50][57][58][67] are extensively used in synchronization purposes by the signal processing community. These bounds are classified into two major categories, namely deterministic bounds (for example: Cramer-Rao bound) and Bayesian bounds (For example: Bayesian Cramer-Rao Bound) [50]. Bayesian bounds are also subdivided into two categories, Ziv-Zakai family, which is derived from a Binary Hypothesis testing problem and the Weiss-Weinstein family, which is derived from a covariance inequality problem[50]. Moreover, these lower bounds are often classified as 'continuous and discrete'[67], 'online and offline'[57][58] and 'unconditional and conditional'[68] based on the context and the availability of the observed information for synchronization purposes.

6. Conclusion

In a *Happy IoT*, service providers' revenue synchronizes to the compliances, inferred from the unreliable sensed data of smart-device users. In this context, we have proposed an unsupervised online Bayesian mechanism *Whiz*, which enables service providers to adapt to the continual change in common interests of Smart-device users. Hence, *Whiz* works as an optimal solution of *online stochastic fact-finding*, such that not only the likelihood of fact (e.g usage-context) is estimated, but also the global performance of likelihood is quantified through online Bayesian filtering, respectively. Extensive experiments on synthetic datasets prove the supremacy of *Whiz* over deterministic *revenue-renewal* model (i.e Tube[45]) and the conventional fact-finders (i.e voting[46], authority and hub[54], Bayesian[43], Likelihood[41] and Cramer-Rao[42] approaches). Accordingly, by utilizing an offline e-mail survey dataset[60] about Smart-device-usage, we have developed prototypes on Android and web-platforms for a context-aware service recommendation scenario for work-going persons. Finally, YouTube[55] and Twitter[56] datasets are trained with *Whiz* to approximate dynamic usage-contexts and to filter social rumor trends, respectively.

7. Appendix

Once common user-groups are initialized, learning model is used by service providers for estimating the likelihood of COI($\theta$) (Lemma. 8) from the incomplete sensed data.

Lemma 8. *The $E$-step and $M$-step of EM are as follows*

*E-Step:* $PlayerA$ computes the log likelihood, where expectation is taken with respect to old $COI(\theta^{t-1})$ and sensed data $D$[69]

$$Q(\theta, \theta^{t-1}) = E[l_c(\theta)|D, \theta^{t-1}] \tag{26}$$

*M-Step:* $PlayerA$ updates $COI(\theta)$, that maximizes the Q-function($Q(\theta, \theta^{t-1})$, derived at E-step), to be used as estimate of $COI(\theta)$ at next iteration.

$$\theta^t = argmax_\theta Q(\theta, \theta^{t-1}) \tag{27}$$

As this unsupervised problem is amenable to general EM, Whiz-EM assists service providers in completing the observed data by iteratively *guessing the latent user-groups* (Lemma. 9)(Lemma. 10) and *then re-estimating the preference (Lemma. 11), discrepancy (Lemma. 12) and mixing strategy (Lemma. 13) of user-groups by using the guessed values as true values*.

**Lemma 9.** *The expected complete data likelihood of Whiz-EM is given by[69]*
$$Q(\theta, \theta^{t-1}) = E[\sum_i \log p(x_i, z_i | \theta)]$$

$$= \sum_i \sum_k \gamma_{ik} \log \pi_k - \frac{1}{2} \sum_i \gamma_{ik}[\log|\Sigma_k| + (x_i - \mu_k)^T \Sigma_k^{-1}(x_i - \mu_k)] \quad (28)$$

*Proof.* $Q(\theta, \theta^{t-1}) = E[\sum_i \log p(x_i, z_i|\theta)]$

$$= \sum_i E[\log[\prod_{k=1}^{K} (\pi_k p(x_i|\theta_k))^{I(z_i=k)}]]$$
$$= \sum_i \sum_k E[I(z_i = k)] \log[\pi_k p(x_i|\theta_k)]$$
$$= \sum_i \sum_k p(z_i = k | x_i, \theta^{t-1}) \log[\pi_k p(x_i|\theta_k)]$$
$$= \sum_i \sum_k \gamma_{ik} \log \pi_k + \sum_i \sum_k \gamma_{ik} \log p(x_i|\theta_k)$$
$$= \sum_i \sum_k \gamma_{ik} \log \pi_k - \frac{1}{2} \sum_i \gamma_{ik}[\log|\Sigma_k| + (x_i - \mu_k)^T \Sigma_k^{-1}(x_i - \mu_k)]$$
□

**Lemma 10.** *The user-group responsibility ($\gamma_{z_{nk}}$) represents the posterior probability of any user-group ($k$) for generating any sensed data point ($x_n$) of any IF(x).*

$$\gamma(z_{nk}) = \frac{\pi_k N(x|\mu_k, \Sigma_k)}{\sum_{j=1}^{K} \pi_j N(x|\mu_j, \Sigma_j)} \quad (29)$$

*Proof.* Let $\pi_k$ be the prior probability of $z_k = 1$, the posterior probability($\gamma(z_{nk})$) for the observed $x$ is calculted with Bayes' theorem

$$\gamma(z_{nk}) = p(z_k = 1|x) = \frac{p(z_k = 1)p(x|z_k = 1)}{\sum_{j=1}^{K} p(z_j=1)p(x|z_j=1)} = \frac{\pi_k N(x|\mu_k, \Sigma_k)}{\sum_{j=1}^{K} \pi_j N(x|\mu_j, \Sigma_j)} \quad (30)$$
□

**Lemma 11.** *The preference($\mu_k$) for k-th user-group is updated by taking the weighted mean of the all sensed data points ($x_n$) of any observed IF(x), where user-group responsibility is used as a weighting factor.*

$$\mu_k = \frac{1}{N_k} \sum_{n=1}^{N} \gamma(z_{nk}) x_n, \quad (31)$$

*where*

$$N_k = \sum_{n=1}^{N} \gamma(z_{nk}) \tag{32}$$

is effective number of sensed data points assigned to user-group k.

*Proof.* Setting the derivative of

$$ln(x|\pi, \mu, \Sigma) \tag{33}$$

with respect to Preference matrix($\mu_k$) to zero yields

$$0 = \sum_{i=1}^{N} \frac{\pi_k N(x_n|\mu_k, \Sigma_k)}{\sum_j \pi_j N(x_n|\mu_j, \Sigma_j)} \Sigma_k^{-1}(x_n - \mu_k) \tag{34}$$

Multiplying by $\Sigma_k$ yields

$$\mu_k = \frac{1}{N_k} \sum_{n=1}^{} \gamma(z_{nk}) x_n, \qquad (35)$$

where

$$N_k = \sum_{n=1}^{} \gamma(z_{nk}) \qquad (36)$$

is effective number of data points assigned to user-group/cluster $k$. □

**Lemma 12.** *The discrepancy matrix($\Sigma_k$) is updated with each sensed data point weighted by user-group responsibility and with denominator given by the effective number of data points, associated with the corresponding component*

$$\Sigma_k = \frac{1}{N_k} \sum_{n=1}^{} \gamma(z_{nk})(x_n - \mu_k)(x_n - \mu_k)^T \qquad (37)$$

*Proof.* Setting the derivative of

$$ln(x|\pi, \mu, \Sigma) \qquad (38)$$

with respect to Discrepancy matrix($\Sigma_k$) to zero, once we obtain

$$\Sigma_k = \frac{1}{N_k} \sum_{n=1}^{} \gamma(z_{nk})(x_n - \mu_k)(x_n - \mu_k)^T \qquad (39)$$

□

**Lemma 13.** *The mixing coefficient of k-th user-group ($\pi_k$) is updated by using the average responsibility, that user-group $k$ takes for generating data ($x_n$) of observed sensed data of IF.*

$$\pi_k = \frac{N_k}{N} \qquad (40)$$

*Proof.* As mixing co-efficients sum to one, by applying a Lagrangian multiplier and maximizing the following

$$lnp(x|\pi, \mu, \Sigma) + \lambda(\sum_{k=1}^{} \pi_k - 1) \qquad (41)$$

we obtain

$$0 = \sum_{n=1} \frac{N_n(x_k | \mu_k, \Sigma)}{\sum_j \pi_j N(x_n | \mu_j, \Sigma_j)} + \lambda \tag{42}$$

Muliplying both sides by $\pi_k$ and summing over $k$, the assumption

$$\sum_{k=1}^{K} \pi_k = 1 \qquad (43)$$

yields

$$\pi_k = \frac{N_k}{N} \qquad (44)$$

□

Online Filtering mode (Definition. 4) is used by service providers to synchronize to the frequent change in common interests among user-groups. Hence, the global quantification (Lemma. 15) of the likelihood of $COI(\theta)$-estimation is achieved with Bayesian online filtering. This is effective, especially, when the uncertainty becomes high and the initial prior knowledge about $COI(\theta)$ suddenly turns to irrelevant, as it evolves over time.

Definition 4. *Online Filtering Mode:* Player A, *as soon as $k$-th sensed data from IF ($x_k$) is received, estimates Player B's $COI(\theta_k)$ by using current and previously sensed data $(x_1^K = [x_1, x_2, \ldots x_k]^T)$ of IF*

Lemma 14. *Cramer-Rao Lower Bound (CRLB) quantifies the local performance of an unbiased and deterministic estimator ($\hat{\theta}(x)$) of the actual $COI(\theta^*)$ by satisfying the following inequality*[50]

$$E_{x|\theta=\theta^*} = [(\hat{\theta}(x) - \theta^*)(\hat{\theta}(x) - \theta^*)^T] > CRLB(\theta^*) \qquad (45)$$

Lemma 15. *Bayesian Cramer-Rao Lower Bound (BCRLB)*[50] *quantifies the global performance of an estimator($\hat{\theta}(x)$) of the actual COI ($\theta^*$) by satisfying the following inequality*

$$E_{x,\theta}\left[(\hat{\theta}(x) - \theta)(\hat{\theta}(x) - \theta)^T\right] \geq BCRLB \qquad (46)$$

*Proof.* BCRLB does not depend on the particular value of $\theta^*$. The BCRB is the inverse of the Bayesian information matrix(BIM), which can be written as

$$B = E_\theta[F(\theta)] + E_\theta[-\triangle_\theta^\theta \log p(\theta)] \qquad (47)$$

where $F(\theta)$ is the Fisher Information Matrix.

BIM's first term is the average information about $\theta$ broght by the observations $x$ and the second term can be regarded as the information available from prior knowledge of $\theta$ that is $p(\theta)$. This allows to take into account the time dependence between COI at different time instants.

□

# References

[1] J. Zheng, D. Simplot-Ryl, C. Bisdikian, H. Mouftah, The internet of things [guest editorial], Communications Magazine, IEEE 49 (11) (2011) 30–31. doi:10.1109/MCOM.2011.6069706.

[2] A. Iera, C. Floerkemeier, J. Mitsugi, G. Morabito, The internet of things [guest editorial], Wireless Communications, IEEE 17 (6) (2010) 8–9. doi:10.1109/MWC.2010.5675772.

[3] N. D. Lane, D. Lymberopoulos, F. Zhao, A. T. Campbell, Hapori: context-based local search for mobile phones using community behavioral modeling and similarity, in: Proceedings of the 12th ACM international conference on Ubiquitous computing, ACM, 2010, pp. 109–118.

[4] A. Madan, M. Cebrian, D. Lazer, A. Pentland, Social sensing for epidemiological behavior change, in: Proceedings of the 12th ACM International Conference on Ubiquitous Computing, Ubicomp '10, ACM, New York, NY, USA, 2010, pp. 291–300. doi:10.1145/1864349.1864394.
URL http://doi.acm.org/10.1145/1864349.1864394

[5] T. Kanda, D. F. Glas, M. Shiomi, H. Ishiguro, N. Hagita, Who will be the customer?: A social robot that anticipates people's behavior from their trajectories, in: Proceedings of the 10th International Conference on Ubiquitous Computing, UbiComp '08, ACM, New York, NY, USA, 2008, pp. 380–389.

[6] S. Hu, L. Su, H. Liu, H. Wang, T. Abdelzaher, Poster abstract: Smartroad: A crowd-sourced traffic regulator detection and identification system, in: Proceedings of the 12th International Conference on Information Processing in Sensor Networks, IPSN '13, ACM, New York, NY, USA, 2013, pp. 331–332.

[7] E. Koukoumidis, L.-S. Peh, M. R. Martonosi, Signalguru: Leveraging mobile phones for collaborative traffic signal schedule advisory, in: Proceedings of the 9th International Conference on Mobile Systems, Applications, and Services, MobiSys '11, ACM, New York, NY, USA, 2011, pp. 127–140.

[8] J. Biagioni, T. Gerlich, T. Merrifield, J. Eriksson, Easytracker: Automatic transit tracking, mapping, and arrival time prediction using smartphones, in: Proceedings of the 9th ACM Conference on Embedded Networked Sensor Systems, SenSys '11, ACM, New York, NY, USA, 2011, pp. 68–81.

[9] A. Thiagarajan, J. Biagioni, T. Gerlich, J. Eriksson, Cooperative transit tracking using smart-phones, in: Proceedings of the 8th ACM Conference on Embedded Networked Sensor Systems, SenSys '10, ACM, New York, NY, USA, 2010, pp. 85–98.

[10] M. Azizyan, I. Constandache, R. Roy Choudhury, Surroundsense: Mobile phone localization via ambience fingerprinting, in: Proceedings of the 15th Annual International Conference on Mobile Computing and Networking, MobiCom '09, ACM, New York, NY, USA, 2009, pp. 261–272.

[11] Y. Chon, N. D. Lane, F. Li, H. Cha, F. Zhao, Automatically characterizing places with opportunistic crowdsensing using smartphones, in: Proceedings of the 2012 ACM Conference on Ubiquitous Computing, UbiComp '12, ACM, New York, NY, USA, 2012, pp. 481–490.

[12] K. K. Rachuri, C. Mascolo, M. Musolesi, P. J. Rentfrow, Sociablesense: Exploring the trade-offs of adaptive sampling and computation offloading for social sensing, in: Proceedings of the 17th Annual International Conference on Mobile Computing and Networking, MobiCom '11, ACM, New York, NY, USA, 2011, pp. 73–84.

[13] L. Deng, L. P. Cox, Livecompare: Grocery bargain hunting through participatory sensing, in: Proceedings of the 10th Workshop on Mobile Computing Systems and Applications, HotMobile '09, ACM, New York, NY, USA, 2009, pp. 4:1–4:6.

[14] A. J. Nicholson, B. D. Noble, Breadcrumbs: Forecasting mobile connectivity, in: Proceedings of the 14th ACM International Conference on Mobile Computing and Networking, MobiCom '08, ACM, New York, NY, USA, 2008, pp. 46–57.

[15] H. Shin, Y. Chon, K. Park, H. Cha, Findingmimo: Tracing a missing mobile phone using daily observations, in: Proceedings of the 9th International Conference on Mobile Systems, Applications, and Services, MobiSys '11, ACM, New York, NY, USA, 2011, pp. 29–42. doi:10.1145/1999995.1999999.
URL http://doi.acm.org/10.1145/1999995.1999999

[16] L. McNamara, C. Mascolo, L. Capra, Media sharing based on colocation prediction in urban transport, in: Proceedings of the 14th ACM International Conference on Mobile Computing and Networking, MobiCom '08, ACM, New York, NY, USA, 2008, pp. 58–69.

[17] P. Zhou, Y. Zheng, M. Li, How long to wait? predicting bus arrival time with mobile phone based participatory sensing, Mobile Computing, IEEE Transactions on 13 (6) (2014) 1228–1241.

[18] M. Demirbas, C. Rudra, A. Rudra, M. Bayir, imap: Indirect measurement of air pollution with cellphones, in: Pervasive Computing and Communications, 2009. PerCom 2009. IEEE International Conference on, 2009, pp. 1–6.

[19] M. Mun, S. Reddy, K. Shilton, N. Yau, J. Burke, D. Estrin, M. Hansen, E. Howard, R. West, P. Boda, Peir, the personal environmental impact report, as a platform for participatory sensing systems research, in: Proceedings of the 7th International Conference on Mobile Systems, Applications, and Services, MobiSys '09, ACM, New York, NY, USA, 2009, pp. 55–68.

[20] R. K. Rana, C. T. Chou, S. S. Kanhere, N. Bulusu, W. Hu, Ear-phone: An end-to-end participatory urban noise mapping system, in: Proceedings of the 9th ACM/IEEE International Conference on Information Processing in Sensor Networks, IPSN '10, ACM, New York, NY, USA, 2010, pp. 105–116.

[21] M. Faulkner, M. Olson, R. Chandy, J. Krause, K. Chandy, A. Krause, The next big one: Detecting earthquakes and other rare events from community-based sensors, in: Information Processing in Sensor Networks (IPSN), 2011 10th International Conference on, 2011, pp. 13–24.

[22] T. Das, P. Mohan, V. N. Padmanabhan, R. Ramjee, A. Sharma, Prism: Platform for remote sensing using smartphones, in: Proceedings of the 8th International Conference on Mobile Systems, Applications, and Services, MobiSys '10, ACM, New York, NY, USA, 2010, pp. 63–76.

[23] A. Liu, J. Bunn, K. Chandy, Sensor networks for the detection and tracking of radiation and other threats in cities, in: Information Processing in Sensor Networks (IPSN), 2011 10th International Conference on, 2011, pp. 1–12.

[24] D. Yang, G. Xue, X. Fang, J. Tang, Crowdsourcing to smartphones: Incentive mechanism design for mobile phone sensing, in: Proceedings of the 18th Annual International Conference on Mobile Computing and Networking, Mobicom '12, ACM, New York, NY, USA, 2012, pp. 173–184.

[25] J.-S. Lee, B. Hoh, Sell your experiences: a market mechanism based incentive for participatory sensing, in: Pervasive Computing and Communications (PerCom), 2010 IEEE International Conference on, 2010, pp. 60–68.

[26] S. Reddy, D. Estrin, M. Hansen, M. Srivastava, Examining micro-payments for participatory sensing data collections, in: Proceedings of the 12th ACM International Conference on Ubiquitous Computing, Ubicomp '10, ACM, New York, NY, USA, 2010, pp. 33–36.

[27] Y. Zhang, M. van der Schaar, Reputation-based incentive protocols in crowdsourcing applications, in: INFOCOM, 2012 Proceedings IEEE, 2012, pp. 2140–2148.

[28] T. Luo, C.-K. Tham, Fairness and social welfare in incentivizing participatory sensing, in: Sensor, Mesh and Ad Hoc Communications and Networks (SECON), 2012 9th Annual IEEE Communications Society Conference on, 2012, pp. 425–433.

[29] M. von Kaenel, P. Sommer, R. Wattenhofer, Ikarus: Large-scale participatory sensing at high altitudes, in: Proceedings of the 12th Workshop on Mobile Computing Systems and Applications, HotMobile '11, ACM, New York, NY, USA, 2011, pp. 63–68.

[30] I. Koutsopoulos, Optimal incentive-driven design of participatory sensing systems, in: INFOCOM, 2013 Proceedings IEEE, 2013, pp. 1402–1410.

[31] B. Faltings, J. Li, R. Jurca, Incentive mechanisms for community sensing, Computers, IEEE Transactions on 63 (1) (2014) 115–128.

[32] H. Liu, S. Hu, W. Zheng, Z. Xie, S. Wang, P. Hui, T. Abdelzaher, Efficient 3g budget utilization in mobile participatory sensing applications, in: INFOCOM, 2013 Proceedings IEEE, 2013, pp. 1411–1419.

[33] J. Rula, F. E. Bustamante, Crowd (soft) control: Moving beyond the opportunistic, in: Proceedings of the Twelfth Workshop on Mobile Computing Systems & Applications, HotMobile '12, ACM, New York, NY, USA, 2012, pp. 3:1–3:6.

[34] X. Bao, M. Gowda, R. Mahajan, R. R. Choudhury, The case for psychological computing, in: Proceedings of the 14th Workshop on Mobile Computing Systems and Applications, HotMobile '13, ACM, New York, NY, USA, 2013, pp. 6:1–6:6.

[35] A. Parate, M. Böhmer, D. Chu, D. Ganesan, B. M. Marlin, Practical prediction and prefetch for

faster access to applications on mobile phones, in: Proceedings of the 2013 ACM International Joint Conference on Pervasive and Ubiquitous Computing, UbiComp '13, ACM, New York, NY, USA, 2013, pp. 275–284.
[36] B. Yan, G. Chen, Appjoy: Personalized mobile application discovery, in: Proceedings of the 9th International Conference on Mobile Systems, Applications, and Services, MobiSys '11, ACM, New York, NY, USA, 2011, pp. 113–126.
[37] X. Bao, S. Fan, A. Varshavsky, K. Li, R. Roy Choudhury, Your reactions suggest you liked the movie: Automatic content rating via reaction sensing, in: Proceedings of the 2013 ACM International Joint Conference on Pervasive and Ubiquitous Computing, UbiComp '13, ACM, New York, NY, USA, 2013, pp. 197–206.
[38] S. Nath, F. X. Lin, L. Ravindranath, J. Padhye, Smartads: bringing contextual ads to mobile apps, in: Proceeding of the 11th annual international conference on Mobile systems, applications, and services, ACM, 2013, pp. 111–124.
[39] K. K. Rachuri, M. Musolesi, C. Mascolo, P. J. Rentfrow, C. Longworth, A. Aucinas, Emotionsense: A mobile phones based adaptive platform for experimental social psychology research, in: Proceedings of the 12th ACM International Conference on Ubiquitous Computing, Ubicomp '10, ACM, New York, NY, USA, 2010, pp. 281–290. doi:10.1145/1864349.1864393.
URL http://doi.acm.org/10.1145/1864349.1864393
[40] H. Barlow, Unsupervised learning, Neural Computation 1 (3) (1989) 295–311.
[41] D. Wang, L. Kaplan, H. Le, T. Abdelzaher, On truth discovery in social sensing: A maximum likelihood estimation approach, in: Proceedings of the 11th international conference on Information Processing in Sensor Networks, ACM, 2012, pp. 233–244.
[42] D. Wang, L. Kaplan, T. Abdelzaher, C. C. Aggarwal, On credibility estimation tradeoffs in assured social sensing, Selected Areas in Communications, IEEE Journal on 31 (6) (2013) 1026–1037.
[43] D. Wang, T. Abdelzaher, H. Ahmadi, J. Pasternack, D. Roth, M. Gupta, J. Han, O. Fatemieh, H. Le, C. C. Aggarwal, On bayesian interpretation of fact-finding in information networks, in: Information Fusion (FUSION), 2011 Proceedings of the 14th International Conference on, IEEE, 2011, pp. 1–8.
[44] K. K. Rachuri, M. Musolesi, C. Mascolo, P. J. Rentfrow, C. Longworth, A. Aucinas, Emotionsense: a mobile phones based adaptive platform for experimental social psychology research, in: Proceedings of the 12th ACM international conference on Ubiquitous computing, pp. 281–290.
[45] S. Ha, S. Sen, C. Joe-Wong, Y. Im, M. Chiang, Tube: time-dependent pricing for mobile data, ACM SIGCOMM Computer Communication Review 42 (4) (2012) 247–258.
[46] A. Gibbard, Manipulation of voting schemes: a general result, Econometrica: journal of the Econometric Society (1973) 587–601.
[47] L. Page, S. Brin, R. Motwani, T. Winograd, The pagerank citation ranking: Bringing order to the web.
[48] J. Kephart, D. Chess, The vision of autonomic computing, Computer 36 (1) (2003) 41–50. doi:10.1109/MC.2003.1160055.
[49] N. Samaan, A. Karmouch, Towards autonomic network management: an analysis of current and future research directions, Communications Surveys & Tutorials, IEEE 11 (3) (2009) 22–36.
[50] A. Renaux, P. Forster, P. Larzabal, C. Richmond, A. Nehorai, A fresh look at the bayesian bounds of the weiss-weinstein family, Signal Processing, IEEE Transactions on 56 (11) (2008) 5334–5352.
[51] R. Kamal, C. S. Hong, S. I. Moon, Whiz-m2m big data learning engine for autonomic happy internet management, KCC 2014 (2014) 1282–1284.
[52] D. Steinley, K-means clustering: a half-century synthesis, British Journal of Mathematical and Statistical Psychology 59 (1) (2006) 1–34.
[53] T. Moon, The expectation-maximization algorithm, Signal Processing Magazine, IEEE 13 (6) (1996) 47–60.
[54] J. M. Kleinberg, Authoritative sources in a hyperlinked environment, Journal of the ACM (JACM) 46 (5) (1999) 604–632.
[55] M. Zeni, D. Miorandi, F. De Pellegrini, Youstatanalyzer: a tool for analysing the dynamics of youtube content popularity, in: Proceedings of the 7th International Conference on Performance Evaluation

Methodologies and Tools, ICST (Institute for Computer Sciences, Social-Informatics and Telecommunications Engineering), 2013, pp. 286–289.
[56] M. De Domenico, A. Lima, P. Mougel, M. Musolesi, The anatomy of a scientific rumor, Scientific reports 3.
[57] S. Bay, C. Herzet, J. Brossier, J. P. Barbot, B. Geller, Analytic and asymptotic analysis of bayesian cramer rao bound for dynamical phase offset estimation, Signal Processing, IEEE Transactions on 56 (1) (2008) 61–70.
[58] H. Hijazi, L. Ros, Analytical analysis of bayesian cramer rao bound for dynamical rayleigh channel complex gains estimation in ofdm system, Signal Processing, IEEE Transactions on 57 (5) (2009) 1889–1900.
[59] X. Yin, J. Han, P. S. Yu, Truth discovery with multiple conflicting information providers on the web, Knowledge and Data Engineering, IEEE Transactions on 20 (6) (2008) 796–808.
[60] R. Kamal, J. Lee, C. Hwang, S. Moon, C. Hong, M. Choi, Psychic: An autonomic inference engine for m2m management in future internet, in: Network Operations and Management Symposium (APNOMS), 2013 15th Asia-Pacific, IEEE, 2013, pp. 1–6.
[61] M. Musolesi, M. Piraccini, A. Corradi, A. Campbell, CRAWDAD data set dartmouth/cenceme (v. 2008-08-13).
[62] J. Lin, S. Amini, J. I. Hong, N. Sadeh, J. Lindqvist, J. Zhang, Expectation and purpose: Understanding users' mental models of mobile app privacy through crowdsourcing, in: Proceedings of the 2012 ACM Conference on Ubiquitous Computing, UbiComp '12, ACM, New York, NY, USA, 2012, pp. 501–510.
[63] E. Miluzzo, C. T. Cornelius, A. Ramaswamy, T. Choudhury, Z. Liu, A. T. Campbell, Darwin phones: the evolution of sensing and inference on mobile phones, in: Proceedings of the 8th international conference on Mobile systems, applications, and services, ACM, 2010, pp. 5–20.
[64] F. Dobrian, V. Sekar, A. Awan, I. Stoica, D. Joseph, A. Ganjam, J. Zhan, H. Zhang, Understanding the impact of video quality on user engagement, SIGCOMM Comput. Commun. Rev. 41 (4) (2011) 362–373.
[65] D. R. Choffnes, F. E. Bustamante, Z. Ge, Crowdsourcing service-level network event monitoring, SIGCOMM Comput. Commun. Rev. 40 (4) (2010) 387–398.
[66] J. Pasternack, D. Roth, Knowing what to believe (when you already know something), in: Proceedings of the 23rd International Conference on Computational Linguistics, Association for Computational Linguistics, 2010, pp. 877–885.
[67] P. Tichavsky, C. Muravchik, A. Nehorai, Posterior cramer-rao bounds for discrete-time nonlinear filtering, Signal Processing, IEEE Transactions on 46 (5) (1998) 1386–1396.
[68] L. Zuo, R. Niu, P. Varshney, Conditional posterior cramer-rao lower bounds for nonlinear recursive filtering, in: Information Fusion, 2009. FUSION '09. 12th International Conference on, 2009, pp. 1528–1535.
[69] K. P. Murphy, Machine learning: a probabilistic perspective, MIT press, 2012.